\PassOptionsToPackage{unicode}{hyperref}
\PassOptionsToPackage{hyphens}{url}
\PassOptionsToPackage{dvipsnames,svgnames,x11names}{xcolor}
\documentclass[
  12pt]{article}

\usepackage{amsmath,amssymb}
\usepackage{tabularx}
\usepackage{siunitx}
\usepackage{iftex}
\ifPDFTeX
  \usepackage[T1]{fontenc}
  \usepackage[utf8]{inputenc}
  \usepackage{textcomp} 
\else 
  \usepackage{unicode-math}
  \defaultfontfeatures{Scale=MatchLowercase}
  \defaultfontfeatures[\rmfamily]{Ligatures=TeX,Scale=1}
\fi
\usepackage{lmodern}
\ifPDFTeX\else  
\fi
\IfFileExists{upquote.sty}{\usepackage{upquote}}{}
\IfFileExists{microtype.sty}{
  \usepackage[]{microtype}
  \UseMicrotypeSet[protrusion]{basicmath} 
}{}
\makeatletter
\@ifundefined{KOMAClassName}{
  \IfFileExists{parskip.sty}{%
    \usepackage{parskip}
  }{
    \setlength{\parindent}{0pt}
    \setlength{\parskip}{6pt plus 2pt minus 1pt}}
}{
  \KOMAoptions{parskip=half}}
\makeatother
\usepackage{xcolor}
\makeatletter
\ifx\paragraph\undefined\else
  \let\oldparagraph\paragraph
  \renewcommand{\paragraph}{
    \@ifstar
      \xxxParagraphStar
      \xxxParagraphNoStar
  }
  \newcommand{\xxxParagraphStar}[1]{\oldparagraph*{#1}\mbox{}}
  \newcommand{\xxxParagraphNoStar}[1]{\oldparagraph{#1}\mbox{}}
\fi
\ifx\subparagraph\undefined\else
  \let\oldsubparagraph\subparagraph
  \renewcommand{\subparagraph}{
    \@ifstar
      \xxxSubParagraphStar
      \xxxSubParagraphNoStar
  }
  \newcommand{\xxxSubParagraphStar}[1]{\oldsubparagraph*{#1}\mbox{}}
  \newcommand{\xxxSubParagraphNoStar}[1]{\oldsubparagraph{#1}\mbox{}}
\fi
\makeatother

\usepackage{longtable,booktabs,array}
\usepackage{calc} 
\usepackage{etoolbox}
\makeatletter
\patchcmd\longtable{\par}{\if@noskipsec\mbox{}\fi\par}{}{}
\makeatother
\IfFileExists{footnotehyper.sty}{\usepackage{footnotehyper}}{\usepackage{footnote}}
\makesavenoteenv{longtable}
\usepackage{graphicx}
\makeatletter
\def\maxwidth{\ifdim\Gin@nat@width>\linewidth\linewidth\else\Gin@nat@width\fi}
\def\maxheight{\ifdim\Gin@nat@height>\textheight\textheight\else\Gin@nat@height\fi}
\makeatother
\setkeys{Gin}{width=\maxwidth,height=\maxheight,keepaspectratio}
\makeatletter
\def\fps@figure{htbp}
\makeatother

\makeatletter
\@ifpackageloaded{caption}{}{\usepackage{caption}}
\AtBeginDocument{%
\ifdefined\contentsname
  \renewcommand*\contentsname{Table of contents}
\else
  \newcommand\contentsname{Table of contents}
\fi
\ifdefined\listfigurename
  \renewcommand*\listfigurename{List of Figures}
\else
  \newcommand\listfigurename{List of Figures}
\fi
\ifdefined\listtablename
  \renewcommand*\listtablename{List of Tables}
\else
  \newcommand\listtablename{List of Tables}
\fi
\ifdefined\figurename
  \renewcommand*\figurename{Figure}
\else
  \newcommand\figurename{Figure}
\fi
\ifdefined\tablename
  \renewcommand*\tablename{Table}
\else
  \newcommand\tablename{Table}
\fi
}
\@ifpackageloaded{float}{}{\usepackage{float}}
\floatstyle{ruled}
\@ifundefined{c@chapter}{\newfloat{codelisting}{h}{lop}}{\newfloat{codelisting}{h}{lop}[chapter]}
\floatname{codelisting}{Listing}

\makeatother
\makeatletter
\@ifpackageloaded{caption}{}{\usepackage{caption}}
\@ifpackageloaded{subcaption}{}{\usepackage{subcaption}}
\makeatother

\ifLuaTeX
  \usepackage{selnolig}  
\fi
\usepackage[]{natbib}
\usepackage{bookmark}

\IfFileExists{xurl.sty}{\usepackage{xurl}}{} 
\hypersetup{
  pdftitle={Title},
  pdfauthor={Author 1; Author 2},
  pdfkeywords={3 to 6 keywords, that do not appear in the title},
  colorlinks=true,
  linkcolor={blue},
  filecolor={Maroon},
  citecolor={Blue},
  urlcolor={Blue},
  pdfcreator={LaTeX via pandoc}}

\newcommand{\anon}{1} 

\usepackage{algorithmic} 
\usepackage{algorithm} 
\usepackage{setspace} 
\usepackage{rotating} 
\usepackage{amsmath} 
\usepackage{amsfonts, amssymb} 
\usepackage{amsthm}
\usepackage{graphicx} 
\usepackage{epstopdf} 
\usepackage{hyperref} 
\usepackage{bm} 
\usepackage{bbm}
\usepackage{multirow} 
\usepackage{latexsym} 
\usepackage{natbib}
\usepackage{rotating} 
\usepackage{titlesec} 
\usepackage{caption} 
\usepackage{mathtools} 
\usepackage{color} 
\usepackage{enumitem} 

\setlist[itemize]{noitemsep, topsep=0pt, leftmargin=0.5cm}

\graphicspath{ {plot/} }
\usepackage{diagbox}
\usepackage{pict2e}
\usepackage{etoolbox}
\usepackage[english]{babel}

\usepackage{cases}
\usepackage{tikz}

\usetikzlibrary{arrows, chains, positioning, quotes, shapes.geometric}

\usepackage{lato}
\usepackage{stmaryrd}

\usepackage{xr}
\makeatletter
\newcommand*{\addFileDependency}[1]{
  \typeout{(#1)}
  \@addtofilelist{#1}
  \IfFileExists{#1}{}{\typeout{No file #1.}}
}
\makeatother

\newcommand*{\myexternaldocument}[1]{%
    \externaldocument{#1}%
    \addFileDependency{#1.tex}%
    \addFileDependency{#1.aux}%
}
\myexternaldocument{DB_JASA_Appendix} 
 
\theoremstyle{definition}
\newtheorem{assumption}{Assumption}

\newtheorem{remark}{Remark}
\newtheorem{example}{Example}

\theoremstyle{theorem}

\newtheorem{theorem}{Theorem}[section]
\newtheorem{lemma}[theorem]{Lemma}

\definecolor{mycolor}{RGB}{0,200,200} 

\newcommand{\indep}{\rotatebox[origin=c]{90}{$\models$} }

\newcommand{\EXP}{\mathbb{E}}
\renewcommand{\Pr}{\mathrm{Pr}}
\newcommand{\VAR}{\text{Var}}
\newcommand{\ind}{\mathbbm{1}}

\DeclareMathOperator*{\argmin}{arg\,min}

\newcommand{\R}{\mathbb{R}}
\newcommand{\E}{\mathbb{E}}

\colorlet{transred}{red!30!white}

\hypersetup{
colorlinks=true,
linkcolor=blue,
filecolor=blue,
urlcolor=blue,
citecolor=blue
}

\newcommand\numeq{\addtocounter{equation}{1}\tag{\theequation}}

\newcommand{\bX}{\mathbf{X}}
\newcommand{\bx}{\mathbf{x}}

\newcommand{\bw}{\mathbf{w}}
\newcommand{\bt}{\mathbf{t}}

\newcommand{\bV}{\mathbf{V}}

\newcommand{\bW}{\mathbf{W}}

\newcommand{\CFD}{\texttt{CFD\hspace{0.1em}}}

\newcommand{\EmpFgp}[1]{F_{\bw,#1,n}}
\newcommand{\Sob}[1]{\mathcal{W}^{#1,2}}

\begin{document}

\def\spacingset#1{\renewcommand{\baselinestretch}%
{#1}\small\normalsize} \spacingset{1}


\if1\anon
{
  \title{\bf Distributional Balancing for Causal Inference: A Unified Framework via Characteristic Function Distance}
  \author{Diptanil Santra$^1$, Guanhua Chen$^2$, Chan Park$^1$\\[0.25cm]
  $^1$: University of Illinois, Urbana-Champaign \\
    $^2$: University of Wisconsin--Madison
    }
    \date{}
  \maketitle
} \fi

\if0\anon
{
  \bigskip
  \bigskip
  \bigskip
\begin{center}
    \LARGE \textbf{Distributional Balancing for Causal Inference:} \\[0.1cm]
    \LARGE \textbf{A Unified Framework via Characteristic} \\ [0.1cm]
    \LARGE \textbf{Function Distance}
\end{center}
  \medskip
} \fi

\bigskip
\begin{abstract} 
Weighting methods are essential tools for estimating causal effects in observational studies, with the goal of balancing pre-treatment covariates across treatment groups. Traditional approaches pursue this objective indirectly, for example, via inverse propensity score weighting or by matching a finite number of covariate moments, and therefore do not guarantee balance of the full joint covariate distributions. Recently, distributional balancing methods have emerged as robust, nonparametric alternatives that directly target alignment of entire covariate distributions, but they lack a unified framework, formal theoretical guarantees, and valid inferential procedures. We introduce a unified framework for nonparametric distributional balancing based on the characteristic function distance (CFD) and show that widely used discrepancy measures, including the maximum mean discrepancy and energy distance, arise as special cases. Our theoretical analysis establishes conditions under which the resulting CFD-based weighting estimator achieves $\sqrt{n}$-consistency. Since the standard bootstrap may fail for this estimator, we propose subsampling as a valid alternative for inference. We further extend our approach to an instrumental variable setting to address potential unmeasured confounding. Finally, we evaluate the performance of our method through simulation studies and a real-world application, where the proposed estimator performs well and exhibits results consistent with our theoretical predictions.

\end{abstract}

\noindent%
\vfill

\newpage
\spacingset{1.8} 

\titlespacing*{\section}{0pt}{0pt}{0pt}
\titlespacing*{\subsection}{0pt}{0pt}{0pt}

\setlength{\abovedisplayskip}{4pt}
\setlength{\belowdisplayskip}{4pt}
\setlength{\abovedisplayshortskip}{2pt}
\setlength{\belowdisplayshortskip}{2pt}
\setlength{\jot}{1pt}
\allowdisplaybreaks

\section{Introduction} \label{sec-Introduction}

A primary objective of causal inference is to estimate causal effects in both randomized experiments and observational studies. In well-executed randomized experiments, treatment assignment is independent of potential outcomes, allowing differences in observed outcomes to be interpreted as causal effects. By contrast, in observational studies, treatment assignment may be confounded with potential outcomes, making naive comparisons of outcomes cannot be interpreted as causal effects and require appropriate adjustments to address confounding bias. A popular approach to addressing confounding in observational studies is weighting, which weights observations to create a pseudo-population in which treated and control groups are more comparable, thereby reducing bias in causal effect estimates; see Section \ref{sec-review} for a review of weighting estimators. 

Among these methods, inverse probability weighting (IPW; \citealp{horvitz1952generalization}) is the most widely used \citep{austin2015moving, hirano2003efficient, lee2010improving, robins2000marginal, robins1995semiparametric}. In general, IPW methods weight each observation by the reciprocal of its propensity score \citep{rosenbaum1983central}, defined as the probability of receiving treatment conditional on observed covariates. However, for their validity, IPW methods typically rely on the correct specification of the propensity score model, as misspecification can introduce substantial bias. Consequently, recent developments in weighting have focused on (near-)balancing covariate moments to improve robustness \citep{Chan2015,cohn2023balancing,hainmueller2012entropy,imai2014covariate,soriano2023interpretable,zubizarreta2015stable}. 
Nevertheless, challenges remain in deciding which moments to balance, and balancing a finite set of moments does not guarantee elimination of confounding beyond those moments. These limitations of IPW and moment-based balancing methods can reduce their reliability and practical applicability in real-world settings; see Sections \ref{sec-simulation} and \ref{sec-real_data} for simulation studies and real-world applications illustrating these issues.

To address these limitations, recent studies have proposed distributional balancing methods \citep{ben2021balancing,bruns2025augmented,chen2024robust,clivio2024towards,HulingGuanhuaIndependence,Huling2024,kallus2018balanced,kallus2020generalized,kong2023covariate,wang2023projected}
which typically proceed in two steps. First, one selects a discrepancy measure to compare full covariate distributions, rather than finite-dimensional summaries such as moments, by comparing the weighted covariate distribution within each treatment group to the unweighted covariate distribution in the overall sample. Second, balancing weights are obtained by minimizing this discrepancy via an optimization procedure (see Section \ref{subsec-CFD based weights} for an example). Unlike IPW and moment-based methods, whose validity hinges on correct specification of the propensity score and the sufficiency of a finite set of moments for removing confounding, distributional balancing directly targets imbalance in the full covariate distributions under weaker assumptions. In practice, such weights can improve empirical performance by achieving balance for arbitrary functions of the covariates without committing to a particular propensity score model or moment conditions that are often unrealistic.

Despite the growing literature, distributional balancing methods remain underdeveloped. First, most existing methods are designed in isolation, tailored to specific distance metrics, and lack a unified framework that accommodates a broad class of metrics. Second, formal theoretical analyses of the resulting weighting estimators are often absent. In particular, it remains unclear whether standard asymptotic properties, such as $\sqrt{n}$-consistency, hold for estimators derived from a given distance metric, and under what conditions these properties are satisfied. Moreover, formal procedures for conducting valid statistical inference on causal effects based on distributional balancing weights remain largely unexplored and often rely naively on bootstrap methods.

In this paper, we develop a novel nonparametric distributional balancing method that addresses multiple limitations of existing approaches. The main contributions of our work can be summarized as follows:

\begin{itemize}

\item[1.] We propose a method based on a characteristic function distance (CFD; \citealp{Ansari2020,CFD1997,CFD2015}), which generalizes several popular distributional balancing methods across a broad class of distance functions. Notably, the construction of the CFD depends on the choice of a density function $\omega$, with certain selections of $\omega$ corresponding directly to established methods. In particular, previous approaches based on the energy distance \citep{Huling2024} and the maximum mean discrepancy \citep{chen2024robust} are special cases of our method when $\omega$ is appropriately specified; see Section \ref{subsec-unification} for a detailed discussion of these relationships. Importantly, our framework is more general than previous methods, as $\omega$ can be chosen flexibly beyond the specific forms required by earlier approaches, provided it satisfies mild regularity conditions.

\item[2.] We clarify how the choice of $\omega$ affects the feasibility of achieving $\sqrt{n}$-consistency for the resulting weighting estimator. Roughly speaking, the tail behavior of $\omega$ determines the smoothness requirements on the outcome regression (i.e., the treatment-specific covariate-outcome relationship). Therefore, we formally establish that distributional balancing with different CFDs necessitates different assumptions on the outcome regression, highlighting that not all choices of $\omega$ guarantee $\sqrt{n}$-consistent estimation for a given dataset; see Section \ref{sec-results} for the details.

\item[3.] We show that bootstrap-based inference may fail to produce reliable results because weighting estimators based on distributional balancing weights may not satisfy the conditions necessary for valid bootstrap inference; see Section \ref{sec-simulation} for empirical evidence from simulation studies. As an alternative, we propose using subsampling \citep{politis1999subsampling}, which remains valid provided that the weighting estimator is $\sqrt{n}$-consistent.

\item[4.] In Section \ref{sec-IV}, we extend our approach to relax the no unmeasured confounding assumption. Specifically, within the instrumental variable (IV) framework of \citet{angrist1996identification}, we show that our distributional balancing method can be directly applied to estimate the local average treatment effect via the Wald estimand in equation \eqref{eq-late wald}. Unlike other regression-based approaches, our weighting estimator based on distributional balancing weights is a one-shot method that does not require estimating multiple nuisance components. We further demonstrate the broader applicability of our method through simulation studies and a real-world application under this IV setting. The results indicate that the CFD-based weighting estimator performs consistently with our theoretical predictions; see Sections \ref{sec-simulation} and \ref{sec-real_data} for details.


\end{itemize}

\section{Setup \& Review}
\subsection{Setup}

Consider an independent and identically distributed (i.i.d.) sample $\{(Y_i, Z_i, \bX_i)\}_{i=1}^n$ of size $n$ generated from a population, where, for unit $i \in \{1,\ldots,n\}$, $Y_i\in\R$ denotes the observed outcome, $Z_i \in \{0, 1\}$ denotes an indicator of whether the unit receives treatment $(Z_i=1)$ or control $(Z_i=0)$, and $\bX_i \in \mathcal{X} \subset \R^d$ denotes a $d$-dimensional vector of pre-treatment covariates. We use $F$ and $\mathcal{X}$ to denote the distribution of $\bX$ and its support, respectively. Let $Y_i(z) \in \R$ be the potential outcome of unit $i \in \{1,\ldots,n\}$, had possibly contrary to fact, the treatment been set to $Z_i=z \in \{0,1\}$. Let $n_1 = \sum_{i=1}^n Z_i$ and $n_0 = \sum_{i=1}^n (1-Z_i)$ denote the number of units received treatment and control, respectively. In addition, we define the empirical cumulative distribution function (ECDF) of the covariates $\{\bX_i\}_{i=1}^n$ as $ F_n(\bx) = \sum_{i=1}^n \mathbb{I}(\bX_i \leq \bx)/n$ and the weighted ECDF of the covariates for units assigned to treatment level $z \in \{0,1\}$, with weights $\bw = (w_1, \ldots, w_n)^\top$, as $\EmpFgp{z}(\bx) = \sum_{i=1}^n w_i\mathbb{I}(\bX_i \leq \bx, Z_i = z)/n_z$. For notational brevity, we omit the subscript $i$ when referring to a generic unit, writing $(Y(z), Y, Z, \bX)$ in place of $(Y_i(z), Y_i, Z_i, \bX_i)$. 

For a vector $\mathbf{v}$, we use $\|\mathbf{v}\| \equiv \|\mathbf{v}\|_2$ to denote its Euclidean ($\ell_2$) norm, and $\|\mathbf{v}\|_1$ and $\|\mathbf{v}\|_\infty$ to denote its $\ell_1$ and $\ell_\infty$ norms, respectively. For a sequence of random variables $\{T_n\}$ and a sequence of positive constants $\{a_n\}$, we write $T_n = O_P(a_n)$ if $T_n / a_n$ is bounded in probability, and $T_n = o_P(a_n)$ if $T_n / a_n \to 0$ in probability. Lastly, let $\Sob{s}(\mathcal{V})$ be the (fractional) Sobolev space consisting of functions whose derivatives up to order $s$ have finite $L_2$ norm over a domain $\mathcal{V}$.

In this article, we focus on the estimation of the average treatment effect (ATE), defined by $
\tau = \E \{Y(1) - Y(0) \} $. Extensions to other causal estimands in an IV setting are discussed in Section \ref{sec-IV}. In order to establish identification of the ATE, we make the following assumptions.

\begin{assumption}[Causal Assumptions] \label{assumption1} 
(i) Stable treatment unit value assumption \citep{Cox1958,Rubin1980}: $Y =  Y(Z)$ almost surely; (ii) Unconfoundedness: $Y(z) \indep Z \, | \,  \bX$ for $z \in \{0,1\}$; (iii) Positivity/Overlap: for some constant $c>0$, the propensity score $e(\bx) = \Pr(Z=1 \, | \, \bX=\bx)$ \citep{rosenbaum1983central} satisfies $e(\bx) \in [c,1-c]$ for all $\bx \in \mathcal{X}$.
\end{assumption}
These assumptions are standard in causal inference; see \citet{HR2020} for a detailed discussion. 
Under Assumption \ref{assumption1}, it is well-known that the ATE can be identified by the following representations \citep{rosenbaum1983central, HR2020}:
\begin{align} \label{eq-ATE}
    \tau =
\E \bigg[ 
\bigg\{
 \frac{Z}{e(\bX)} - \frac{1-Z}{1-e(\bX)}
 \bigg\} Y
\bigg]
=
\E \big\{ \mu_1(\bX) - \mu_0(\bX) \big\} \ ,
\end{align} 
where $e(\bX)$ is the propensity score and $\mu_z(\bX) \equiv \E ( Y | Z=z, \bX )$ is the outcome regression.

\subsection{Review: Weighting Estimators} \label{sec-review}
 


The first representation of \eqref{eq-ATE} serves as the key basis for weighting approaches, as it demonstrates that applying appropriate weights to treated and control units can eliminate confounding bias and allow for recovering causal effects. This principle---eliminating covariate imbalance via weighting---is the underlying idea behind these approaches. Formally, let $\bw=(w_1,\ldots,w_n)^\top$ denote non-negative weights assigned to each unit. Given these weights, a weighting estimator of the ATE can be expressed as:
\begin{align} \label{eq-WATE}
\widehat{\tau}_\bw
= \frac{1}{n_1} \sum_{i=1}^n w_i Z_i Y_i - \frac{1}{n_0} \sum_{i=1}^n w_i (1-Z_i) Y_i \ .    
\end{align}
Following \citet{Huling2024}, the error of $\widehat{\tau}_{\bw}$ can be decomposed as:
\begin{align}
    \widehat{\tau}_\bw - \tau =
    & \int_{\mathcal{X}} \mu_1(\bx)d(\EmpFgp{1} - F_n)(\bx) - \int_{\mathcal{X}} \mu_0(\bx)d(\EmpFgp{0} - F_n)(\bx)\label{H&M 3}\\
    & \quad - \int_{\mathcal{X}} \{ \mu_1(\bx) - \mu_0(\bx) \} \, d(F - F_n) (\bx)
    + \sum_{i=1}^n w_i \varepsilon_i \bigg( \frac{Z_i}{n_1} - \frac{1-Z_i}{n_0} \bigg),
    \label{H&M 5}
\end{align}
where $\varepsilon_i \equiv Y_i - \mu_{Z_i}(\bX_i)$, satisfying $\E (\varepsilon_i | Z_i,\bX_i) = 0$. Note that the terms in \eqref{H&M 5} converge to zero as the sample size grows. Consequently, any asymptotic bias of $\widehat{\tau}_{\bw}$ as $n \rightarrow \infty$ arises from \eqref{H&M 3}. In other words, the main source of potential bias in $\widehat{\tau}_{\bw}$ is the discrepancy between $\EmpFgp{z}$ and $F_n$, i.e., the weighted ECDFs and the unweighted ECDF. This underscores why the central aim in constructing weights is to minimize, ideally eliminate, imbalances in the covariate distributions of treated and control units relative to the marginal distribution.




We now provide a brief overview of how balancing weights can be constructed. The first approach is inverse probability weighting (IPW; \citealp{horvitz1952generalization,hirano2003efficient}), which constructs weights based on the estimated propensity score $\widehat{e}$ and yields the following IPW estimator of the ATE:
\begin{align} \label{eq-IPW} 
    w_{i,\text{IPW}}
    =
    \frac{1}{n}
    \bigg\{
    \frac{n_1 Z_i}{\widehat{e}(\bX_i) }
    +
    \frac{n_0 (1-Z_i)}{1-\widehat{e}(\bX_i)}
    \bigg\}
    \ , \quad 
    \widehat{\tau}_{\text{IPW}} = \frac{1}{n}\sum_{i=1}^n \left\{\frac{Z_i}{\widehat{e}(\bX_i)} - \frac{(1-Z_i)}{1 - \widehat{e}(\bX_i)}\right\} Y_i \ .
\end{align}
The propensity score may be estimated using parametric methods (e.g., logistic regression) or nonparametric methods (e.g., machine learning). When the true propensity score is known, it can replace $\widehat{e}$ in the above expressions. A normalized version, known as the H\'ajek estimator \citep{hajek1971comment}, rescales the weights so that they sum to one within treatment groups, i.e., $w_{i,\text{Hajek}} \propto w_{i,\text{IPW}}$ with $\sum_{i=1}^{n} \mathbb{I}(Z_i=z) w_{i,\text{Hajek}} = n_z$ for $z \in \{0,1\}$. The H\'ajek estimator often improves finite-sample stability compared to the IPW estimator.

The second approach is the covariate balancing propensity score (CBPS; \citealp{imai2014covariate}), in which the propensity score is estimated by solving a set of moment equations that directly enforce covariate balance for a user-specified function $h(\cdot)$, often taken as the identity function. Specifically, the CBPS estimate $\widehat{e}_B$ is desiged to satisfy:
\[
0 \;=\; \frac{1}{n} \sum_{i=1}^n 
\left\{\frac{Z_i}{\widehat{e}_B(\bX_i)} - \frac{1 - Z_i}{1 - \widehat{e}_B(\bX_i)}\right\} h(\bX_i) \ .
\]  
By construction, the CBPS estimator $\widehat{e}_B$ is guaranteed to achieve covariate balance with respect to the specified basis function $h$, a property that the standard propensity score estimator $\widehat{e}$ in \eqref{eq-IPW} may not generally possess. The resulting CBPS weighting estimator has the same form as \eqref{eq-IPW}, with $\widehat{e}_B$ replacing $\widehat{e}$.

The third approach is entropy balancing \citep{hainmueller2012entropy, Chan2015}, where the weights are obtained by solving a constrained optimization problem, say:
\begin{align*}
\min_{\bw} \sum_{i=1}^n h(w_i)
\quad 
\text{ subject to }    
\quad 
\bigg\{
    \begin{array}{l}
    \sum_{i=1}^n w_i c_{ri}(\bX_i) = m_r \quad \text{with } r \in \{ 1, \dots, R \} 
    \\[-0.2cm]
    \sum_{i=1}^n w_i \mathbb{I}(Z_i=z) = n_z \ , \ z \in \{ 0 , 1 \}
    \ , \quad 
    \bw \geq 0 
    \end{array}
\end{align*} 
where $h(w_i)$ is a distance metric between $w_i$ and  reference weights (typically the uniform weights) and $c_{ri}(\bX_i) = m_r$ describes a set of $R$ balance constraints imposed on the covariate moments. By construction, entropy balancing guarantees that, after weighting, the specified moments of the covariates are exactly matched within each treatment group.

Lastly, a recent development in constructing weights is distributional balancing \citep{ben2021balancing, bruns2025augmented,chen2024robust,clivio2024towards,HulingGuanhuaIndependence,Huling2024,kallus2018balanced,kallus2020generalized,kong2023covariate,wang2023projected}, which aims to achieve balance across the full covariate distributions. In this approach, weights are obtained by minimizing a distance between two weighted empirical covariate distributions; see Section \ref{subsec-CFD based weights} for a formal mathematical characterization of these weights. Examples of such distances include maximum mean discrepancy (MMD; \citealp{Gretton2012}), energy distance (ED; \citealp{szekely2004testing}), the Kolmogorov-Smirnov statistic, $f$-divergences, and the Wasserstein distance. Unlike the aforementioned approaches, distributional balancing does not require specification of a propensity score model or a priori selection of moments to be balanced. 
Consequently, it is a nonparametric method that is robust to bias from model misspecification and can achieve covariate balance not only for selected moments but across the entire covariate distributions. As detailed in the remainder of this manuscript, our approach falls within the distributional balancing framework based on the characteristic function distance (CFD;  \citealp{Ansari2020,CFD1997,CFD2015}).


\section{Method: Unified Distributional Balancing}

\subsection{Characteristic Function Distance (CFD)}

To introduce our approach, we first provide a brief review of the CFD. The CFD between two distributions $P$ and $Q$ with respect to a 
nonnegative density $\omega: \R^d \rightarrow \R$ is defined as
\begin{align}\label{def:CFD}
    \CFD^2_\omega(P,Q)
    \;=\;
    \int_{\R^d} \big| \varphi_P(\bt) - \varphi_Q(\bt)\big|^2 \,\omega(\bt) \, d\bt,
\end{align}

where $\varphi_P(\bt)=\E_{\bV\sim P}\{ \exp ({\sqrt{-1}\,\bt^\top \bV} ) \}$ is the characteristic function of $P$ (and similarly for $\varphi_Q$). The density function $\omega$ is assumed to satisfy certain properties, namely:
\begin{assumption} \label{assumption-omega}
$\omega$ is nonnegative, finite, and integrable, with support equal to $\R^d$.
\end{assumption}
Under Assumption \ref{assumption-omega}, the corresponding CFD defines a proper metric between distributions, as formalized in the following lemma, which is a direct consequence of Theorem 9 of \citet{sriperumbudur2010hilbert}:
\begin{lemma}\label{lemma-characterization of CFD}
    Under Assumption \ref{assumption-omega}, we have $\CFD_\omega^2(P,Q)=0 $ if and only if $P=Q$.
\end{lemma}

Under Assumption \ref{assumption-omega}, we may, without loss of generality, normalize $\omega$ so that it integrates to one, making it a probability density function on $\R^d$. While the finiteness and integrability requirements could be relaxed for certain cases (see Example \ref{example-Energy Distance} below), we retain them here for simplicity and to avoid unnecessary technical complications.

The CFD admits equivalent representations that highlight its connection to reproducing kernel Hilbert spaces (RKHS) and Fourier analysis. Given a density function $\omega$ satisfying Assumption \ref{assumption-omega}, consider the translation-invariant kernel
\begin{align} \label{eq-kernel} 
    k_{\omega}(\bx,\bx') =\int_{\R^d} e^{\sqrt{-1}\,\bt^\top (\bx - \bx')} \, \omega(\bt) \, d\bt \ , 
\end{align}
which is well-defined and positive definite by Bochner’s theorem \citep{bochner2005harmonic}, since $\omega$ acts as the spectral density of the kernel. In particular, $k_{\omega}$ is the inverse Fourier transform of $\omega$, and $\omega$ is the Fourier transform of $k_{\omega}$. Then, the CFD between $P$ and $Q$ can be expressed in kernel form as
\begin{align}\label{eq:CFD-kernel-form}
    \CFD^2_\omega(P,Q)
    &= \E_{\bV,\bV' \sim P, \ \bW,\bW' \sim Q}\big\{ k_{\omega}(\bV,\bV') + k_{\omega}(\bW,\bW') - 2 k_{\omega}(\bV,\bW)\big\},
\end{align}
where $(\bV,\bV',\bW,\bW')$ are mutually independent with $\bV,\bV' \sim P$ and $\bW,\bW' \sim Q$. Furthermore, $\CFD_\omega^2$ admits a variational characterization as an MMD: 
\begin{align}\label{eq:CFD-dual}  
    \CFD^2_\omega(P,Q)
    = \left[\sup_{f \in \mathcal{H}_\omega:\,\|f\|_{\mathcal{H}_\omega}\le 1} \big[ \E_{\bV \sim P, \bW \sim Q}\{f(\bV) - f(\bW)\} \big]\right]^2,
\end{align}  
where $\mathcal{H}_\omega$ is the RKHS associated with $k_\omega$, and $\| \cdot \|_{\mathcal{H}_{\omega}}$ is the corresponding RKHS norm. In other words, $\CFD_\omega^2(P,Q)$ is precisely the squared MMD between $P$ and $Q$ with respect to the kernel $k_{\omega}$. See Appendix \ref{appendix-Equivalence of CFD forms} for details on the derivations of \eqref{eq:CFD-kernel-form} and \eqref{eq:CFD-dual}. In fact, the RKHS $\mathcal{H}_{\omega}$ is often characterized as a Sobolev space whose smoothness is determined by the choice of $\omega$; see Examples \ref{example-Gaussian density}-\ref{example-Energy Distance} below.

We now focus on the case where the two distributions of interest are the observed data, i.e., $\widehat{P}$ and $\widehat{Q}$ are the ECDFs of $\{\bV_i\}_{i=1}^{n} \sim P$ and $\{\bW_j\}_{i=1}^{m}\sim Q$, respectively. The goal is to evaluate the CFD between $\widehat{P}$ and $\widehat{Q}$. If the kernel function $k_{\omega}$ in \eqref{eq-kernel} admits a closed-form representation, the CFD can be expressed in terms of the gram matrix as 
\begin{align}\label{eq:CFD-gram}
\!\!\!\CFD^2_\omega(\widehat{P},\widehat{Q})
& \! = \!
\frac{1}{n^2}\sum_{i=1}^n\sum_{j=1}^n k_\omega(\bV_i, \bV_j)
\! + \! \frac{1}{m^2}\sum_{i=1}^m\sum_{j=1}^m k_\omega(\bW_i, \bW_j)
\! - \! \frac{2}{nm}\sum_{i=1}^n\sum_{j=1}^m k_\omega(\bV_i, \bW_j).
\end{align}

However, obtaining a closed-form kernel may be infeasible when the density $\omega$ has a complex structure; see Examples \ref{example-Separable-density} and \ref{example-Isotropic-density}. In such cases, the kernel \eqref{eq-kernel} can be approximated via Monte Carlo. Specifically, for a sufficiently large $L$, let $\{ \mathbf{T}_\ell \}_{\ell=1}^{L} $  be i.i.d. draws from a distribution with the probability density function proportional to $\omega$. Then, $k_{\omega}$ can be approximated by $\tilde{k}_{\omega}$, which serves as the empirical analogue of \eqref{eq-kernel}:
\begin{align*}
    \tilde{k}_{\omega}(\bx,\bx')
    & =\frac{1}{L}\sum_{\ell=1}^L e^{\sqrt{-1} \mathbf{T}_\ell^\top(\bx-\bx')}.
\numeq
\label{eq:approx-CFD} 
\end{align*}
This approximation can be computed efficiently using the addition formula for trigonometric functions. Moreover, if $\omega$ is symmetric about zero, i.e., $\omega(\bt)=\omega(-\bt)$, then \eqref{eq:approx-CFD} reduces to an expression that involves only the real part; see Appendix \ref{approximation to cos} for details.


\subsection{CFD-based Balancing Weights}\label{subsec-CFD based weights}

As established in the preceding analysis, the primary source of potential bias in the weighting estimator $\widehat{\tau}_\bw$ arises from the term in \eqref{H&M 3}, which reflects the discrepancy between the weighted empirical distributions for the treated and control groups, $\EmpFgp{1}$ and $\EmpFgp{0}$, and the unweighted empirical distribution of the full sample, $F_n$. To reduce this bias, we seek weights $\bw$ such that the weighted distributions $\EmpFgp{z}$ become as similar as possible to the target distribution $F_n$ in terms of the CFD. This objective can be formalized as an optimization problem: minimize the sum of the CFDs between each treatment-group empirical distribution and the full-sample distribution. If this sum attains zero, the weighted empirical distributions are exactly equal to the target distribution from Lemma \ref{lemma-characterization of CFD}, i.e., perfectly balanced. Otherwise, the solution yields weights that render the distributions as similar as possible under the CFD metric.

Following this motivation, we define the CFD balancing weights as the solution to:
\begin{align}
    \bw_{\CFD}
    \in 
    & \ \argmin_{{\bw}} 
    \big\{ {\CFD}_{\omega}^2(\EmpFgp{1} , F_{n}) + {\CFD}_{\omega}^2(\EmpFgp{0} , F_{n}) +
    \lambda^2 \|\bw\|_2^2 \big\}
    \nonumber
    \\
    &
     \
    \text{subject to}
    \quad 
    \sum_{i=1}^{n} w_i Z_i = n_1 \ , \quad 
    \sum_{i=1}^{n} w_i (1-Z_i) = n_0 \ , \quad 
    \bw \geq 0
    \ .
    \label{CFD balancing weights}
\end{align}
Here, $\lambda \geq 0$ is a regularization parameter introduced to prevent overfitting, which is required to decay at a certain rate with respect to $n$; see Assumption \ref{assumption-lambda} for details. The constraints $\sum_{i=1}^n w_i Z_i = n_1$ and $\sum_{i=1}^n w_i (1-Z_i) = n_0$ are included to normalize the weights within each treatment group. Their primary purpose is to ensure that the total ``mass'' or effective sample size of the weighted groups remains equal to that of the original, unweighted groups. This normalization is crucial for two reasons. First, it provides scale preservation, preventing the weights from arbitrarily inflating or deflating the influence of a group, which would distort the final estimate. Second, it leads to stable estimation, making the corresponding weighting estimators (e.g., $n_1^{-1}\sum w_i Y_i Z_i$) directly comparable in form and scale to their simple unweighted counterparts. This practice is analogous to the normalization of weights in the H\'ajek estimator reviewed in Section \ref{sec-review}, which generally offers superior finite-sample properties relative to the IPW estimator.

\begin{remark} \label{subsec-3 way balancing}
The formulation in \eqref{CFD balancing weights} aims to find weights that make each treatment arm resemble the full sample but does not directly enforce similarity between the treatment and control groups. As a result, distributional discrepancies between the treated and control groups may remain, particularly in finite samples or when the two groups differ substantially in size or composition. To address this issue, one can adopt a ``three-way'' balancing objective that adds a direct discrepancy term between the two weighted groups. Following \citet{Chan2015} and \citet{Huling2024}, we consider
\begin{align}
    \bw_{3,\CFD}
    \in 
    & \ \argmin_{{\bw}} 
    \big\{ {\CFD}_{\omega}^2(\EmpFgp{1} , F_{n}) + {\CFD}_{\omega}^2(\EmpFgp{0} , F_{n}) +
    {\CFD}_{\omega}^2(\EmpFgp{1} , \EmpFgp{0})+
    \lambda^2 \|\bw\|_2^2 \big\}
    \nonumber
    \\
    &
     \
    \text{subject to}
    \quad 
    \sum_{i=1}^{n} w_i Z_i = n_1 \ , \quad 
    \sum_{i=1}^{n} w_i (1-Z_i) = n_0 \ , \quad 
    \bw \geq 0
    \ .
    \label{CFD balancing weights 3-way}
\end{align}
The additional term ${\CFD}_{\omega}^2(\EmpFgp{1}, \EmpFgp{0})$ directly penalizes dissimilarity between the two groups' empirical distributions under the CFD metric, aiming to achieve covariate balance in which the treated group, control group, and full sample are jointly aligned.
\end{remark}

Now, we use \eqref{eq:CFD-gram} as a basis for solving the optimization tasks \eqref{CFD balancing weights} and \eqref{CFD balancing weights 3-way}. Specifically, we take $\widehat{P}$ to be the weighted treated sample $\EmpFgp{1}$ and $\widehat{Q}$ to be the unweighted full sample $F_n$. The weighted adaptation of the formula is:
\begin{align}
    \CFD^2_\omega(\EmpFgp{1},F_{n})
    &
    =
    \underbrace{\frac{1}{n_1^2}\sum_{i,j} w_i w_j Z_i Z_j k(\bX_i, \bX_j)}_{
    \substack{
    \text{Weighted treated group vs.} 
    \\         
    \text{Weighted treated group} 
    }}
    + \underbrace{\frac{1}{n^2}\sum_{i,j} k(\bX_i, \bX_j)}_{
    \substack{
    \text{Full sample vs.} 
    \\         
    \text{Full sample} 
    }}
    - \underbrace{\frac{2}{n_1 n}\sum_{i,j} w_i Z_i k(\bX_i, \bX_j)}_    {
    \substack{
    \text{Weighted treated group vs.} 
    \\         
    \text{Full sample} 
    }}
    \nonumber
    \\
    &
    =
    \frac{1}{n_1^2} \bw^\top D_{1} K D_{1} \bw 
    + \frac{1}{n^2}\mathbf{1}^\top K \mathbf{1} 
    - \frac{2}{n_1 n} \bw^\top D_{1} K \mathbf{1} \ .
    \label{CFD-Q-1}
\end{align}
In the first line, $\sum_{i,j}$ is  shorthand $\sum_{i=1}^{n}\sum_{j=1}^{n}$. 
In the second line, $K \in \R^{n \times n}$ is the gram matrix of which $(i,j)$th entry is $k_{\omega}(\bX_i,\bX_j)$, $D_{1}= \text{diag}(Z_1,\ldots,Z_n) \in \R^{n \times n}$ is the diagonal matrix indicating the treated units, and $\mathbf{1} =(1,\ldots,1)^\top \in \R^{n}$ is a vector of ones. Similarly, taking $(\widehat{P},\widehat{Q})= (\EmpFgp{0},F_n)$ and $(\widehat{P},\widehat{Q})= (\EmpFgp{1},\EmpFgp{0})$ yields
\begin{align}
    &
    \CFD^2_\omega(\EmpFgp{0},F_{n})
    =
    \frac{1}{n_0^2} \bw^\top D_{0} K D_{0} \bw 
    + \frac{1}{n^2}\mathbf{1}^\top K \mathbf{1}
    - \frac{2}{n_0 n} \bw^\top D_{0} K \mathbf{1} ,
    \label{CFD-Q-0}
    \\ 
    &
    \CFD^2_\omega(\EmpFgp{1}, \EmpFgp{0}) = \frac{1}{n_1^2} \bw^\top D_{1} K D_{1} \bw  - \frac{2}{n_1 n_0}\bw^\top D_{1} K D_{0}\bw + \frac{1}{n_0^2} \bw^\top D_{0} K D_{0} \bw , 
    \label{CFD-Q-0-1}
\end{align}
where $D_{0} = \text{diag}(1-Z_1,\ldots,1-Z_n) \in \R^{n \times n}$ is the diagonal matrix indicating the control units.   Substituting \eqref{CFD-Q-1}-\eqref{CFD-Q-0-1} into \eqref{CFD balancing weights 3-way}, we obtain  
\begin{align} \label{eq-QP}
&
\bw_{\CFD}
\in
\argmin_{\bw \in \R^n} 
\big( \bw^\top \mathcal{Q} \bw + \mathbf{q}^\top \bw \big)
\ \
    \text{s.t.}
    \ \
    \sum_{i=1}^{n} w_i Z_i = n_1, \
    \sum_{i=1}^{n} w_i (1-Z_i) = n_0 , \
    \bw \geq 0 ,
\\
&
\mathcal{Q}
=
\frac{D_{1} K D_{1}}{n_1^2}  + \frac{D_{0} K D_{0}}{n_0^2} -\frac{D_{1} K D_{0}}{n_1n_0}  + \lambda^2 I_n
\ , \quad 
\mathbf{q} = -\left( \frac{D_{1} K \mathbf{1} }{n_1 n} + \frac{D_{0} K \mathbf{1}}{n_0 n}  \right) \ .
\nonumber
\end{align}
A similar formulation can be derived for the case of \eqref{CFD balancing weights}. Equation \eqref{eq-QP} is a standard constrained quadratic programming (QP) task with a unique global solution, as the objective function is strictly convex for $\lambda > 0$ \citep{boyd2004convex}. This problem can be solved using general-purpose QP solvers; for instance, we used an R library \texttt{optiSolve} \citep{optiSolve} in the simulation and data analysis; see Appendix \ref{sec-QP} on details of how this is implemented. Of note, when the closed-form representation of the CFD is not available, the approximated kernel in \eqref{eq:approx-CFD} can be used in \eqref{CFD-Q-1}-\eqref{eq-QP}.

Finally, using the estimated weights $\bw_{\CFD} = (w_{1,\CFD}, \ldots, w_{n,\CFD})^\top$, we construct the weighting estimator of the ATE from \eqref{eq-WATE}, denoted by $\widehat{\tau}_{\CFD}$, as follows:
\begin{align} \label{eq-tau cfd}
    \widehat{\tau}_{\CFD}
    = \frac{1}{n_1} \sum_{i=1}^n w_{i,\CFD} Z_i Y_i - \frac{1}{n_0} \sum_{i=1}^n w_{i,\CFD} (1-Z_i) Y_i \ . 
\end{align}

\subsection{Unification: Examples of Our Framework}\label{subsec-unification}

Many distributional distances can be expressed in the CFD form of \eqref{def:CFD} with a suitable choice of the density function $\omega$. Consequently, our approach provides a unified framework for existing distributional balancing methods. Several examples are discussed below.

\begin{example}[Gaussian Density] \label{example-Gaussian density}
A prominent example is the MMD with a Gaussian kernel, which arises when $\omega(\bt)$ is a multivariate Gaussian density, specifically $\omega_{\text{G}}(\bt) \propto \exp(-\gamma^2 \|\bt\|_2^2/4)$ for a bandwidth parameter $\gamma > 0$. The inverse Fourier transform of this Gaussian density yields the Gaussian kernel, denoted by $k_{\text{G}}(\bx, \bx') = \exp(-\|\bx-\bx'\|_2^2 / \gamma^2)$. The corresponding RKHS $\mathcal{H}_{\text{G}}$ is a collection of functions that are exceptionally smooth; specifically, $\mathcal{H}_{\text{G}}$ is contained in the intersection of infinitely many Sobolev spaces:
$\mathcal{H}_{\text{G}} \subset \Sob{s}(\mathcal{X})$ for any $s>0$ \citep[Corollary 4.36]{SVM2008}, meaning that every function in $\mathcal{H}_{\text{G}}$ is infinitely differentiable.
\end{example}

\begin{example}[Separable Density] \label{example-Separable-density}
Another common choice for the density function is a separable one, which can be written as a product of one-dimensional densities. The $\ell_1$-Laplacian kernel provides a key instance of this case. It corresponds to selecting $\omega_{\text{L}}(\bt) \propto \prod_{j=1}^d (1 + \gamma^2 t_j^2)^{-1}$, which is a product of one-dimensional Cauchy densities. The resulting kernel is the $d$-dimensional $\ell_1$-Laplacian kernel, $k_{\text{L}}(\bx, \bx') = \exp(-\|\bx-\bx'\|_1 / \gamma) = \prod_{j=1}^d \exp(-|x_j - x_j'| / \gamma)$, which itself is a product of one-dimensional Laplacian kernels.

The one-dimensional kernel $\exp(-|x_j - x_j'|/\gamma)$ has the one-dimensional Sobolev space $\Sob{1}(\mathcal{X}_j)$ as its native space \citep{wendland2004scattered} where $\mathcal{X}_j$ is the support of $X_{j}$. Since $k_{\text{L}}$ is a tensor product of these one-dimensional kernels, its native space is the tensor product Sobolev space, i.e.,  $\mathcal{H}_{\text{L}} = \bigotimes_{j=1}^d \Sob{1}(\mathcal{X}_j)$ \citep[Theorem 13]{berlinet2011reproducing}.

The above Cauchy density can be replaced with generic one-dimensional densities. For instance, we can select $\omega_{\text{S},j}(t_j)$ to be a scaled density of the Student's t-distribution with $(2s_j-1)$ degrees of freedom, which corresponds to
$
\omega_{\text{S},j}(t_j) \propto (1 + \gamma^2 t_j^2)^{-s_j}.
$
For a certain $s_j$, the corresponding one-dimensional kernel $k_{\text{S},j}$ has a known closed form as a Mat\'ern kernel \citep{Rasmussen2006Gaussian}. The native space for the one-dimensional kernel $k_{\text{S},j}$ is the Sobolev space $\Sob{s_j}(\mathcal{X}_j)$, and consequently, the native space for the full kernel $k_{\text{S}}$ is $\mathcal{H}_{\text{S}} = \bigotimes_{j=1}^d \Sob{s_j}(\mathcal{X}_j)$. 
Notably, the Laplacian kernel $k_{\text{L}}$ and the Gaussian kernel $k_{\text{G}}$ in Example \ref{example-Gaussian density} correspond to the special cases $s_j = 1$ and $s_j=\infty$, respectively, for all $j$. Therefore, specifying the Laplacian and Gaussian kernels can be seen as two extreme cases, while the general case with $s_j \in (1, \infty)$ represents an intermediate one between them.

In fact, $\omega_{\text{S},j}$ does not need to be exactly proportional to the t-distribution density. Specifically, suppose there exist positive constants $c_1$ and $c_2$ such that $
c_1 (1 + \gamma^2 t_j^2)^{-s_j} 
\leq 
\omega_{\text{S},j}(t_j) 
\leq 
c_2
(1 + \gamma^2 t_j^2)^{-s_j} 
$. This relaxation allows the density $\omega_{\text{S},j}(t_j)$ to be complex, provided that its tail behavior resembles that of a t-distribution. In such cases, obtaining a closed-form expression for the coordinate-wise and full kernels, $k_{\text{S},j}$ and $k_{\text{S}}$, may be challenging. Nevertheless, these kernels can be effectively approximated using Monte Carlo integration via \eqref{eq:approx-CFD}. Importantly, despite this relaxation, the native space associated with the full kernel $k_{\text{S}}$ remains $\mathcal{H}_{\text{S}} = \otimes_{j=1}^d \Sob{s_j}(\mathcal{X}_j)$ \citep[Corollary 10.13]{wendland2004scattered}.

\end{example}

\begin{example}[Isotropic Density]\label{example-Isotropic-density}
Our framework also includes isotropic (i.e., rotationally invariant) densities. For example, consider the density $\omega_{\text{M}}(\bt) \propto (1 + \gamma^2 \|\bt\|_2^2)^{-s}$, where $s$ is a smoothness parameter with $s > d/2$. Unlike the previous example, this density cannot be decomposed into a product of one-dimensional functions. Consequently, the corresponding kernel $k_{\text{M}}$ is a $d$-dimensional Mat\'ern kernel rather than the product of one-dimensional Mat\'ern kernel; see Appendix \ref{Spectral Densities for Different Kernels} for the closed-form representation of it. The resulting native space $\mathcal{H}_{\text{M}}$ is a $d$-dimensional Sobolev space $\Sob{s}(\mathcal{X})$ \citep[Chapter 10]{wendland2004scattered}, which differs from $\mathcal{H}_{\text{S}}$ in the previous example. As in the previous example, this condition can be relaxed: specifically, if there exist positive constants $c_1$ and $c_2$ such that $c_1 (1 + \gamma^2 \|\bt\|_2^2)^{-s} \leq \omega_{\text{M}}(\bt) \leq c_2 (1 + \gamma^2 \|\bt\|_2^2)^{-s}$, then the resulting native space remains $\mathcal{H}_{\text{M}} = \Sob{s}(\mathcal{X})$ \citep[Corollary 10.13]{wendland2004scattered}.
\end{example}

\begin{example}[Energy Distance]\label{example-Energy Distance}
While primarily focused on proper densities $\omega$ satisfying Assumption \ref{assumption-omega}, our framework can also accommodate certain ``improper densities,'' with the ED serving as a key example.  For the ED, the associated density is $\omega_{\text{ED}}(\bt) \propto \|\bt\|_2^{-(d+1)}$, which does not satisfy Assumption \ref{assumption-omega}. Consequently, the corresponding kernel cannot be obtained via the standard inverse Fourier transform in \eqref{eq-kernel}. Nevertheless, the inverse Fourier transform exists in a distributional sense, yielding the energy distance kernel $k_{\text{ED}}(\bx, \bx') = -\|\bx - \bx'\|_2$ \citep{Szekely2013}. Note that this kernel is not positive definite, as Bochner’s theorem does not apply. However, following \citet{mak2018support}, the associated native space $\mathcal{H}_{\text{ED}}$ can still be characterized and is closely related to Sobolev spaces. Specifically, the Sobolev space $\Sob{(d+1)/2}$ is embedded in $\mathcal{H}_{\text{ED}}$, namely $\Sob{(d+1)/2} \subset \mathcal{H}_{\text{ED}}$. Moreover, when the dimension $d$ is odd, the spaces are exactly equal: $\Sob{(d+1)/2} = \mathcal{H}_{\text{ED}}$. This provides a unified mathematical representation of the ED as a type of CFD, since the formulations in \eqref{def:CFD}, \eqref{eq:CFD-kernel-form}, and \eqref{eq:CFD-dual} remain valid for the ED, even though it is not a standard CFD.

\end{example}

\begin{remark}[Exception: Wasserstein distance]\label{example-Wasserstein}

Another popular discrepancy metric is the Wasserstein $p$-distance ($W_p$). While the discussion can be extended to general $p$, we focus on $p=1$ as a motivating example. Note that the $W_1$ distance is represented as $W_1(P,Q) = \sup_{f \in \mathcal{F}} [ \E_{\bV \sim P, \bW \sim Q}\{f(\bV) - f(\bW)\} ]$, where $\mathcal{F}$ is the class of 1-Lipschitz functions. This representation shares similarity as the CFD in \eqref{eq:CFD-dual}, except that the function classes defining the distances differ. However, \citet{modeste2024characterization} showed that the Wasserstein distance cannot be represented in the CFD form of \eqref{def:CFD}. Consequently, the Wasserstein distance is a notable exception that our method does not unify. 

Nonetheless, this exception does not diminish the practical advantages of our approach. The native space for the $W_1$ distance (i.e., 1-Lipschitz functions) is significantly more complex than the Sobolev spaces discussed above, as demonstrated by a formal comparison of their covering numbers in Appendix \ref{Comparison between different Function Classes}. A direct consequence of this complexity is that the $W_1$ distance suffers from the curse of dimensionality; specifically, it is well-known that $W_1(F_n,F) = O_P(n^{-1/d})$ for $d \geq 3$ \citep[page 137]{vanHandel2014}, leading to substantially slower convergence than the $\sqrt{n}$-rate. Therefore, even when the corresponding balancing weights are well-calibrated and the outcome regressions ($\mu_z$) are sufficiently smooth, using the $W_1$ distance may yield a weighting estimator that fails to achieve $\sqrt{n}$-consistency. In contrast, our unified CFD-based approach produces a $\sqrt{n}$-consistent weighting estimator as long as $\mu_z$ exhibits an appropriate degree of smoothness relative to the specified density $\omega$. More specifically, if $\mu_z$ belongs to the native space corresponding to $\omega$, $\sqrt{n}$-consistency can be established; see Theorem \ref{theorem ATE} for details.

\end{remark}

\section{Theory}\label{sec-results}

\subsection{$\sqrt{n}$-Consistency of the Weighting Estimator}

We now present the statistical properties of the weighting estimator, where the weights are obtained from the CFD balancing problem in \eqref{CFD balancing weights 3-way}. We note that the statistical properties of the weighting estimator based on \eqref{CFD balancing weights} can be established analogously. We begin by specifying the required assumptions.


\begin{assumption}[Compact Covariate Support] \label{assumption2}
The covariate support $\mathcal{X}$ is compact.
\end{assumption}

Assumption \ref{assumption2} ensures that uniform boundedness and convergence of integrals involving kernels, distance metrics, and characteristic functions. 

\begin{assumption}[Smoothness of the Outcome Regression]\label{assumption-mu}
$\mu_z(\cdot) \in \mathcal{H}_\omega$ for $z \in \{0,1\}$.
\end{assumption}

Assumption \ref{assumption-mu} requires the true outcome regressions $\mu_z$ to lie within the native space $\mathcal{H}_\omega$. This is a critical smoothness assumption because properties of $\mathcal{H}_\omega$ are entirely determined by the choice of $\omega$. This directly connects to one of our contributions, which highlights that the tail behavior of $\omega$ dictates the smoothness of the functions contained in its native space. For instance, a rapidly decaying $\omega$ (e.g., Gaussian) induces a native space consisting of exceptionally smooth functions, thereby imposing a very strong condition on $\mu_z$. In contrast, a heavy-tailed $\omega$ (e.g., Laplacian) generates a native space that contains less smooth functions, corresponding to a weaker, less restrictive assumption. Thus, the choice of $\omega$ implicitly specifies the smoothness required of the true outcome model. 

\begin{assumption}[Boundeness of the Outcome] \label{assumption-sigma}
For $z \in \{0,1\}$, 
    $\text{Var}\{ \mu_z(\bX) \} < \infty$ and $\sigma_z^2(\bX) \equiv \VAR(Y \, | \, Z=z, \bX) $ is bounded over $\bX \in \mathcal{X}$.
\end{assumption}

Assumption \ref{assumption-sigma} imposes conditions on the outcome distribution; these conditions are satisfied when $Y$ is uniformly bounded.

\begin{assumption}[Stable Weights] \label{assumption-bound on weights}
    The balancing weights $\bw_{\CFD} = (w_{1,\CFD},\ldots,w_{n,\CFD})^\top$ in \eqref{CFD balancing weights 3-way} satisfy $\|\bw_{\CFD}\|_{\infty} \leq Cn^{1/3}$ for some constant $C > 0$ independent of $n$.
\end{assumption}

Assumption \ref{assumption-bound on weights} states that the estimated balancing weights are not ``too large'', specifically that no single weight grows faster than an $n^{1/3}$-rate. Several weighting-based methods for covariate balancing rely on this assumption \citep{wong2018kernel,athey2018approximate,Huling2024}. It prevents a pathological scenario where one or a few data points receive enormous weight (e.g., weights on the order of $n$), which would make the estimator highly unstable by depending on just a few observations.

\begin{assumption}[Regularization Rate] \label{assumption-lambda}
The regularization parameter in \eqref{CFD balancing weights 3-way} is chosen with a rate of $\lambda = o(n^{-1})$.
\end{assumption}

Assumption \ref{assumption-lambda} guarantees that the penalty term in \eqref{CFD balancing weights 3-way}, $\lambda^2 \|\bw\|_2^2$, becomes asymptotically negligible relative to the CFD terms.

Under these assumptions, the CFD-based weighting estimator in \eqref{eq-tau cfd} is $\sqrt{n}$-consistent:
\begin{theorem}\label{theorem ATE}
    Suppose that Assumptions \ref{assumption1}-\ref{assumption-lambda} are satisfied. Then, the proposed weighting estimator $\widehat{\tau}_{\CFD}$ in \eqref{eq-tau cfd} with balancing weights $\bw_{\CFD}$ from \eqref{CFD balancing weights 3-way} is $\sqrt{n}$-consistent, i.e., $[\E_{Y,Z,\bX} \{ (\widehat{\tau}_{\CFD} - \tau)^2\}]^{1/2} = O(n^{-1/2})$.

\end{theorem}

Leveraging the result in Theorem \ref{theorem ATE}, we can construct confidence intervals for inference; see Section \ref{sec:subsampling} for details. We emphasize that Assumption \ref{assumption-omega} is a sufficient condition used to simplify the presentation; in some cases, $\sqrt{n}$-consistency can still hold even when it is violated.

\begin{remark}\label{relaxing assumption A4}
    Although the density $\omega$ can be chosen to satisfy Assumption \ref{assumption-omega}, it is also of interest to consider choices that do not. In fact, Theorem \ref{theorem ATE} may still hold even when Assumption \ref{assumption-omega} is violated. For example, consider the ED density from Example \ref{example-Energy Distance}, $\omega_{\text{ED}}(\bt) = \|\bt\|_2^{-(d+1)}$. Although this choice clearly violates Assumption \ref{assumption-omega}, the corresponding weighting estimator remains $\sqrt{n}$-consistent; see \citet{Huling2024} for a formal proof of this result. Additional intuition is provided in Remark \ref{remark: relaxation of omega for ED} in the Appendix.
    

\end{remark}

\begin{remark}\label{augmentation}
A consequence of semiparametric efficiency theory \citep{BKRW1998} is that any regular and asymptotically linear (RAL) estimator of the ATE in the nonparametric model, denoted by $\widehat{\tau}_{\text{RAL}}$, admits the following unique asymptotic representation:
\begin{align} \label{eq-asymp lin}   
&
\sqrt{n}(\widehat{\tau}_{\text{RAL}} - \tau)
\nonumber
\\
&
= \frac{1}{\sqrt{n}}\sum_{i=1}^n 
\underbrace{
\bigg[
\frac{(2Z_i -1) \{Y_i-\mu_{Z_i}(\bX_i)\} }{Z_i e(\bX_i) + (1-Z_i) \{1-e(\bX_i)\}} + \mu_1(\bX_i) - \mu_0(\bX_i) - \tau
\bigg]}_{
\texttt{IF}(Y_i,Z_i,\bX_i)} + o_P(1),
\end{align} 
Here, $\texttt{IF}(\cdot)$ is an influence function for the ATE in the nonparametric model \citep{hahn1998role}, which is unique. While Theorem \ref{theorem ATE} establishes the $\sqrt{n}$-consistency of our estimator $\widehat{\tau}_{\CFD}$, this result alone does not guarantee that $\widehat{\tau}_{\CFD}$ admits the asymptotic linear representation in \eqref{eq-asymp lin}, i.e., $\widehat{\tau}_{\CFD}$ may not be RAL. This is because (i) our estimator relies on the weights $\bw_{\CFD}$ obtained from the constrained QP \eqref{eq-QP}, which are subject to boundary constraints that typically induce non-regularity; and (ii) the outcome regression is not explicitly incorporated, a crucial component for ensuring regularity.

Although $\widehat{\tau}_{\CFD}$ is not guaranteed to be RAL, one could in principle obtain an RAL estimator by strengthening assumptions or adding modeling components. For example, imposing additional smoothness conditions on the propensity score can yield efficiency gains in the spirit of \citet{hirano2003efficient}. Alternatively, one could combine the CFD weights with an outcome model-based augmentation term, as in \citet{hirshberg2021augmented}. We leave such efficiency-oriented extensions to future work.

The lack of a general RAL guarantee for $\widehat{\tau}_{\CFD}$ is not necessarily a drawback for our goals. Rather, our method is complementary to RAL-based approaches. Constructing RAL estimators typically requires either (i) stronger structural assumptions (e.g., smooth propensity scores); (ii) fitting one or more nuisance regressions, which can be burdensome when multiple outcomes or multiple nuisance functions arise (see Section~\ref{sec-IV}); or (iii) cross-fitting with sample splitting to establish theoretical guarantees \citep{DDML2018}. In contrast, our approach estimates balancing weights without using outcome information or specifying propensity score models, and the same set of weights can be reused to estimate causal effects for many outcomes. In addition, our method does not require cross-fitting, allowing characterization of its asymptotic properties based on in-sample estimation. For inference, while RAL estimators leverage asymptotic normality, our procedure relies on methods that are valid under $\sqrt{n}$-consistency without requiring asymptotic normality. This distinction requires careful discussion, which we provide in the next Section.

\end{remark}

\subsection{Inference via Subsampling}\label{sec:subsampling}

Bootstrap \citep{efron1994introduction} is a widely used method and is often regarded as a readily accessible tool for conducting inference in practice. Despite its popularity, a theoretically important  but less frequently discussed point is that its validity generally relies on the estimator being regular, i.e., it can fail for non-regular estimators \citep{shao1994bootstrap,abadie2008failure,andrews2000inconsistency,chakraborty2010inference}. As discussed in Remark \ref{augmentation}, our estimator may not be regular, implying that the bootstrap procedure may not yield valid statistical inference. Indeed, we observe in our simulations that the standard bootstrap can fail to provide valid confidence intervals; see Section \ref{sec-simulation}. 

As a valid alternative, we consider subsampling. Specifically,  \citet{politis1994large} showed that subsampling remains valid for a wide class of estimators, including non-regular ones, as long as the estimator's rate of convergence is known. Since we established the $\sqrt{n}$-consistency of the proposed weighting estimator in Theorem \ref{theorem ATE}, subsampling provides a valid method for constructing confidence intervals. Nonetheless, subsampling often produces conservative inference, even for large sample sizes, a phenomenon commonly observed in practice. For theoretical explanations of this conservativeness, see \citet{andrews2010asymptotic}. A practical challenge in implementing subsampling is the choice of the subsample size. To address this, we propose a data-driven method for selecting the subsample size following the volatility-based procedure of \citet{politis1999subsampling}. The full details of this selection algorithm are provided in Appendix \ref{appendix-subsampling}. 

\section{Extension to an Instrumental Variable Setting}\label{sec-IV}

The proposed method relies on the no unmeasured confounding assumption, namely Assumption \ref{assumption1}(ii), but it can be extended to account for unmeasured confounding. Specifically, we consider the IV framework of \citet{angrist1996identification}. Let $\bX \in \mathcal{X}$ denote the vector of observed pre-treatment covariates, $Z \in \{0,1\}$ the binary instrument, $A \in \{0,1\}$ the binary indicator of treatment receipt, where $A=1$ indicates that a unit receives the treatment and $A=0$ otherwise, and $Y \in \R$ the outcome variable. Let $A(z)$ denote the potential treatment status under $Z=z$, $Y(z,a)$ the potential outcome under $(Z=z,A=a)$, and $Y(z)=Y(z,A(z))$ the potential outcome under $Z = z$ with $A$ set to the natural treatment receipt. We impose the following IV assumptions, as detailed in \citet{angrist1996identification} and \citet{Levis2024}:
\begin{assumption}[Instrumental Variable] \label{assumption IV} (i) Consistency: $Y = Y(Z)$ and $A=A(Z)$ almost surely; (ii) Exclusion Restriction: $Y(z=1,a) = Y(z=0,a)$ almost surely for $a \in \{0,1\}$; (iii) Unconfoundedness: $ \{ A(z), Y(z) \} \indep Z \, | \,  \bX$ for $z \in \{0,1\}$; (iv) Relevance: $A \not\!\!\indep Z \, |\,  \bX$; (v) Monotonicity/No Defiers: $A(1) \geq A(0)$ almost surely. 
\end{assumption}

In this context, a primary causal estimand of interest is the local average treatment effect (LATE), defined as $\tau_{\text{LATE}}=\E\{Y(a=1) - Y(a=0) \mid A(z=1) > A(z=0)\}$, assuming $\Pr\{ A(z=1) > A(z=0) \} > 0$ to ensure that the LATE is well-defined. Under Assumptions \ref{assumption1}(iii) and \ref{assumption IV}, the LATE is identified as:
\begin{align}
    \tau_{\text{LATE}} = 
    \frac{\E \{ \mu_1(\bX) - \mu_0(\bX) \}}{ \E \{ \nu_1(\bX) - \nu_0(\bX) \} } \ , \ \mu_z(\bX) = \EXP(Y \, | \, Z=z,\bX) \ , \ \nu_z(\bX) = \EXP(A \, | \, Z=z, \bX) \ .
    \label{eq-late wald}
\end{align}
See Appendix \ref{appendix-LATE} for details on replicating this result. Note that both the numerator and denominator have the same form as \eqref{eq-ATE}; specifically, they correspond to the ATEs of $Z$ on $Y$ and $Z$ on $A$, respectively. Consequently, one can estimate these ATEs using weighting estimators, with weights derived from the proposed distributional balancing approach. Specifically, let $\bw_{\CFD}$ denote the balancing weights defined in \eqref{CFD balancing weights 3-way}. The resulting weighting estimator of the LATE is then given by:
\begin{align}
    \widehat{\tau}_{\text{LATE},\CFD} = \frac{ \frac{1}{n_1}\sum_{i=1}^n Z_i w_{i,\CFD} Y_i - \frac{1}{n_0}\sum_{i=1}^n (1-Z_i) w_{i,\CFD} Y_i }{ \frac{1}{n_1}\sum_{i=1}^n Z_i w_{i,\CFD} A_i - \frac{1}{n_0}\sum_{i=1}^n (1-Z_i) w_{i,\CFD} A_i }.\label{estimator of LATE} 
\end{align}
We note that $\widehat{\tau}_{\text{LATE},\CFD}$ is a one-shot approach, as both the numerator and denominator rely solely on the weights. In contrast, regression-based methods require separately estimating $\mu_z$ and $\nu_z$. This illustrates the point made in Remark \ref{augmentation} where our approach avoids the need to estimate multiple nuisance functions, whereas other methods do not.

Under assumptions analogous to those in Theorem \ref{theorem ATE}, along with Assumption \ref{assumption IV}, it can be shown that $\widehat{\tau}_{\text{LATE},\CFD}$ is $\sqrt{n}$-consistent for $\tau_{\text{LATE}}$. The details are provided below; we omit their interpretation, as they parallel those corresponding to the earlier assumptions and theorem.


\begin{assumption}[Smoothness of Nuisance Functions]\label{assumption-mu modified}
For $z \in \{0,1\}$, both the outcome regression $\mu_z(\cdot)$ and the treatment regression $\nu_z(\cdot)$ belong to $\mathcal{H}_\omega$.
\end{assumption}




\begin{theorem}\label{theorem LATE}

    Suppose that Assumptions \ref{assumption1}(iii), \ref{assumption-omega}-\ref{assumption2}, \ref{assumption-sigma}-\ref{assumption-mu modified} are satisfied. Then, $\widehat{\tau}_{\text{LATE},\CFD}$ is $\sqrt{n}$-consistent, i.e., $[\E \{ (\widehat{\tau}_{\text{LATE},\CFD} - \tau_{\text{LATE}})^2\}]^{1/2} = O(n^{-1/2})$.

\end{theorem}

\section{Simulation}\label{sec-simulation} 

We conduct a simulation study to evaluate the finite-sample performance of the proposed CFD-based weighting estimators. To illustrate our method’s performance beyond the standard no unmeasured confounding scenario, we consider the following data-generating process under the IV setting outlined in Section \ref{sec-IV}. First, we fix the sample size to $n \in \{100, 200, 400, 800, 1600\}$. For each subject $i \in \{1,\ldots,n\}$, we generate a $10$-dimensional pre-treatment covariate $\bX_i=(X_{i1},\ldots,X_{i10})^\top$, whose components are independent standard normal random variables, i.e., $\bX_i \sim N(\mathbf{0}, I_{10})$. We also generate a variable $U_i \sim N(0,1)$, independently from $\bX_i$. We treat $\bX_i$ and $U_i$ as measured and unmeasured confounders, respectively.

The binary instrument $Z_i$ is generated from $Z_i \sim \text{Ber}(e(\bX_i))$, where the propensity score $e$ is specified according to one of two models: (linear) $e_{\text{Linear}} (\bX_i)
    =
    \text{expit}
    \big( 0.15 \sum_{j=1}^{10} X_{ij} \big)$; or (nonlinear) $e_{\text{Nonlinear}} (\bX_i)
    =
    \text{expit}
    \big( 0.25 \sum_{j=1}^{5} | X_{ij} | + 0.25 \sum_{j=6}^{10}  X_{ij}  -2.25\big)$, where $\text{expit}(v) = \exp(v)/\{1+\exp(v)\}$. The potential treatment receipts under each instrument level, $A_i(0)$ and $A_i(1)$, are generated as $A_{i}(0) = \ind ( L_{1i} - 0.5 | U_i | \geq 0 ) $ and $A_{i}(1) = \ind ( L_{1i} + 0.5 | U_i | \geq 0 )$ where $L_{1i} = - 0.01 - 0.5 \sum_{j=1}^{5} X_{ij} + 0.5 \sum_{j=6}^{10} X_{ij}$. This specification ensures that the monotonicity assumption holds, i.e., $A_i(1) \ge A_i(0)$ for all $i$.  The potential outcomes are generated as $Y_i(a) = a( 1 + L_{2i}) - 0.5 L_{2i} + \varepsilon_{ai}$ for $a \in \{0,1\}$ where $L_{2i} = \sum_{j=1}^{5}X_{ij} - 0.5 \sum_{j=6}^{10} X_{ij} - U_i$ and noise terms $\varepsilon_{ai}$ are independently drawn from $N(0,1)$. The observed treatment and outcome are given by $A_{i} = A_{i}(Z_i)$ and $Y_i = Y_i(A_i)$. In the analysis, we use $(Y_i, A_i, Z_i, \bX_i)$, while $U_i$ is omitted as it serves as an unmeasured confounder.

We estimate the LATE using our proposed CFD-based weighting estimators with five different specifications of the density function $\omega$: (i) \textit{Gaussian}: $\omega_{\text{G}}$ from Example \ref{example-Gaussian density}, which yields the Gaussian kernel; (ii) \textit{$\ell_1$-Laplacian}: $\omega_{\text{L}}$ from Example \ref{example-Separable-density}, producing the $\ell_1$-Laplacian kernel; (iii) \textit{t$_5$}:  $\omega_{\text{S}}$ from Example \ref{example-Separable-density}, where $\omega_{\text{S},j}$ is proportional to a t-distribution with 5 degrees of freedom; (iv) \textit{Mat\'ern}: $\omega_{\text{M}}$ from Example \ref{example-Isotropic-density} with $s=7.5$, producing the Mat\'ern kernel; and (v) \textit{energy}: $\omega_{\text{ED}}$ from Example \ref{example-Energy Distance}, corresponding to the energy distance. Except for (iii), the CFD is evaluated using the closed-form representation of the kernel $k_{\omega}$, whereas for (iii), it is approximated using \eqref{eq:approx-CFD} with $L=10^4$ Monte Carlo repetitions. For (i)-(iv), the density requires selecting bandwidth hyperparameters $\gamma$ (see Examples \ref{example-Gaussian density}-\ref{example-Isotropic-density}), which we choose using the median heuristic \citep{garreau2018largesampleanalysismedian}. For all proposed estimators, we use the three-way balancing in Remark \ref{subsec-3 way balancing}, and the regularization parameter in \eqref{CFD balancing weights 3-way} is set to $\lambda = n^{-2}$, which satisfies Assumption \ref{assumption-lambda}. In addition, we construct 95\% confidence intervals using both subsampling and bootstrap methods for inference. As competing methods, we also consider weighting estimators based on weights derived from the parametric IPW and CBPS approaches described in Section \ref{sec-review}. For the IPW estimator, $e(\cdot)$ is estimated using logistic regression of $Z$ on $\bX$ with main-effect terms only; for CBPS, we use the implementation in the \texttt{WeightIt} R package \citep{WeightIt}, which estimates propensity score weights under the same logistic specification while targeting balance of covariate means. Further details on the median heuristic, the construction of confidence intervals, and the implementation of the IPW and CBPS estimators are provided in Appendices \ref{appendix-median-heuristic}-\ref{implementation of the IPW and CBPS estimators}. We repeat the simulation 500 times, computing the bias and empirical standard error (ESE) for all seven methods. For inference, we report the empirical coverage rates of the 95\% subsampling and bootstrap confidence intervals for our weighting estimators (i)-(v), and those of bootstrap confidence intervals for the IPW and CBPS estimators.

Table \ref{tab:simulation} presents a summary of the simulation results. We begin by examining the bias and ESE of the competing estimators. In both the linear and non-linear propensity score model scenarios, the bias and ESE of our weighting estimators all decrease steadily toward zero as the sample size $n$ increases, and the decrease appears to follow an $n^{-1/2}$-rate as demonstrated in Appendix~\ref{sec:validation of root_n consistency}. This result is consistent with Theorem \ref{theorem ATE}. In contrast, the IPW and CBPS methods show a critical vulnerability to model misspecification. While these estimators appear consistent under the linear propensity score setting, they fail in the non-linear propensity score scenario, as their bias remains large and persistent regardless of sample size. This failure is expected, given the severe misspecification of their underlying parametric propensity score models.

Next, we compare the subsampling and bootstrap inference procedures. Across all scenarios and for all five weighting estimators, the subsampling confidence intervals attain the nominal 95\% coverage level. This confirms that subsampling provides valid, albeit slightly conservative, inference for our estimators. In contrast, the bootstrap confidence intervals perform poorly, particularly for (ii) the $\ell_1$-Laplacian and (v) energy distance cases. This finding empirically supports our earlier concern regarding the failure of the standard bootstrap for the weighting estimator, as discussed in Section \ref{sec:subsampling}. While the subsampling intervals are consistently wider than the bootstrap intervals, this moderate increase in width is reasonable and reflects the method’s conservative nature, which guards against undercoverage. In comparison, the bootstrap intervals may fail dramatically, providing misleadingly narrow intervals with substantial coverage deficits. Consequently, we recommend subsampling as the preferred method for inference. The results clearly demonstrate the statistical properties and robustness to model misspecification of the proposed CFD-based weighting estimators, while also confirming subsampling as a reliable inferential procedure.

{\begin{table}[!ht]
\renewcommand{\arraystretch}{1}
\scriptsize
\centering
\setlength{\tabcolsep}{4pt} 

\begin{tabular}{|cc|c|cccccc|cccccc|}
\hline
\multicolumn{2}{|c|}{\multirow{4}{*}{Method}} & \multirow{4}{*}{$N$} & \multicolumn{12}{c|}{Data-generating process}                                                                                                                                                                                                                                                                                                                                                          \\ \cline{4-15} 
\multicolumn{2}{|c|}{}                        &                      & \multicolumn{6}{c|}{Linear}                                                                                                                                                                        & \multicolumn{6}{c|}{Nonlinear}                                                                                                                                                \\ \cline{4-15} 
\multicolumn{2}{|c|}{}                        &                      & \multicolumn{1}{c|}{\multirow{2}{*}{Bias}} & \multicolumn{1}{c|}{\multirow{2}{*}{ESE}} & \multicolumn{2}{c|}{Coverage}                       & \multicolumn{2}{c|}{Length}                         & \multicolumn{1}{c|}{\multirow{2}{*}{Bias}} & \multicolumn{1}{c|}{\multirow{2}{*}{ESE}} & \multicolumn{2}{c|}{Coverage}                       & \multicolumn{2}{c|}{Length}    \\ \cline{6-9} \cline{12-15} 
\multicolumn{2}{|c|}{}                        &                      & \multicolumn{1}{c|}{}                      & \multicolumn{1}{c|}{}                     & \multicolumn{1}{c|}{SS} & \multicolumn{1}{c|}{Boot} & \multicolumn{1}{c|}{SS} & \multicolumn{1}{c|}{Boot} & \multicolumn{1}{c|}{}                      & \multicolumn{1}{c|}{}                     & \multicolumn{1}{c|}{SS} & \multicolumn{1}{c|}{Boot} & \multicolumn{1}{c|}{SS} & Boot \\ \hline
\multicolumn{1}{|c|}{\multirow{25}{*}{\begin{tabular}[c]{@{}c@{}}Distributional\\ Balancing\end{tabular}}} & \multirow{5}{*}{\begin{tabular}[c]{@{}c@{}}Example \ref{example-Gaussian density}\\ Gaussian\end{tabular}} & 100 & \multicolumn{1}{c|}{0.9} & \multicolumn{1}{c|}{62.6} & \multicolumn{1}{c|}{97.8} & \multicolumn{1}{c|}{94.0} & \multicolumn{1}{c|}{4.5} & 2.6 & \multicolumn{1}{c|}{-5.7} & \multicolumn{1}{c|}{67.6} & \multicolumn{1}{c|}{95.8} & \multicolumn{1}{c|}{92.0} & \multicolumn{1}{c|}{5.0} & 2.8 \\ \cline{3-15} 
\multicolumn{1}{|c|}{} & & 200 & \multicolumn{1}{c|}{0.9} & \multicolumn{1}{c|}{43.4} & \multicolumn{1}{c|}{98.0} & \multicolumn{1}{c|}{95.4} & \multicolumn{1}{c|}{2.4} & 1.6 & \multicolumn{1}{c|}{1.2} & \multicolumn{1}{c|}{44.3} & \multicolumn{1}{c|}{98.0} & \multicolumn{1}{c|}{94.0} & \multicolumn{1}{c|}{2.5} & 1.6 \\ \cline{3-15} 
\multicolumn{1}{|c|}{} & & 400 & \multicolumn{1}{c|}{1.2} & \multicolumn{1}{c|}{30.5} & \multicolumn{1}{c|}{98.0} & \multicolumn{1}{c|}{96.4} & \multicolumn{1}{c|}{1.5} & 1.1 & \multicolumn{1}{c|}{1.2} & \multicolumn{1}{c|}{30.2} & \multicolumn{1}{c|}{98.2} & \multicolumn{1}{c|}{95.8} & \multicolumn{1}{c|}{1.5} & 1.1 \\ \cline{3-15} 
\multicolumn{1}{|c|}{} & & 800 & \multicolumn{1}{c|}{-1.3} & \multicolumn{1}{c|}{20.4} & \multicolumn{1}{c|}{98.6} & \multicolumn{1}{c|}{98.2} & \multicolumn{1}{c|}{1.0} & 0.8 & \multicolumn{1}{c|}{-0.2} & \multicolumn{1}{c|}{22.7} & \multicolumn{1}{c|}{97.8} & \multicolumn{1}{c|}{96.2} & \multicolumn{1}{c|}{1.0} & 0.8 \\ \cline{3-15} 
\multicolumn{1}{|c|}{} & & 1600 & \multicolumn{1}{c|}{-0.4} & \multicolumn{1}{c|}{16.0} & \multicolumn{1}{c|}{96.4} & \multicolumn{1}{c|}{96.6} & \multicolumn{1}{c|}{0.6} & 0.6 & \multicolumn{1}{c|}{-0.5} & \multicolumn{1}{c|}{15.2} & \multicolumn{1}{c|}{98.6} & \multicolumn{1}{c|}{99.2} & \multicolumn{1}{c|}{0.7} & 0.6 \\ \cline{2-15} 
\multicolumn{1}{|c|}{} & \multirow{5}{*}{\begin{tabular}[c]{@{}c@{}}Example \ref{example-Separable-density}\\ $\ell_1$-Laplacian\end{tabular}} & 100 & \multicolumn{1}{c|}{1.9} & \multicolumn{1}{c|}{65.5} & \multicolumn{1}{c|}{97.8} & \multicolumn{1}{c|}{90.8} & \multicolumn{1}{c|}{5.0} & 2.5 & \multicolumn{1}{c|}{-13.1} & \multicolumn{1}{c|}{72.5} & \multicolumn{1}{c|}{94.8} & \multicolumn{1}{c|}{88.0} & \multicolumn{1}{c|}{5.7} & 2.8 \\ \cline{3-15} 
\multicolumn{1}{|c|}{} & & 200 & \multicolumn{1}{c|}{0.4} & \multicolumn{1}{c|}{42.2} & \multicolumn{1}{c|}{97.2} & \multicolumn{1}{c|}{91.8} & \multicolumn{1}{c|}{2.6} & 1.6 & \multicolumn{1}{c|}{-6.2} & \multicolumn{1}{c|}{46.2} & \multicolumn{1}{c|}{96.8} & \multicolumn{1}{c|}{89.8} & \multicolumn{1}{c|}{2.8} & 1.6 \\ \cline{3-15} 
\multicolumn{1}{|c|}{} & & 400 & \multicolumn{1}{c|}{0.9} & \multicolumn{1}{c|}{28.1} & \multicolumn{1}{c|}{98.0} & \multicolumn{1}{c|}{91.4} & \multicolumn{1}{c|}{1.5} & 1.0 & \multicolumn{1}{c|}{-1.5} & \multicolumn{1}{c|}{30.5} & \multicolumn{1}{c|}{95.8} & \multicolumn{1}{c|}{91.4} & \multicolumn{1}{c|}{1.6} & 1.0 \\ \cline{3-15} 
\multicolumn{1}{|c|}{} & & 800 & \multicolumn{1}{c|}{-0.7} & \multicolumn{1}{c|}{19.7} & \multicolumn{1}{c|}{97.8} & \multicolumn{1}{c|}{89.4} & \multicolumn{1}{c|}{1.0} & 0.7 & \multicolumn{1}{c|}{-1.4} & \multicolumn{1}{c|}{20.0} & \multicolumn{1}{c|}{97.2} & \multicolumn{1}{c|}{90.2} & \multicolumn{1}{c|}{1.0} & 0.7 \\ \cline{3-15} 
\multicolumn{1}{|c|}{} & & 1600 & \multicolumn{1}{c|}{-0.4} & \multicolumn{1}{c|}{13.2} & \multicolumn{1}{c|}{98.2} & \multicolumn{1}{c|}{90.4} & \multicolumn{1}{c|}{0.7} & 0.4 & \multicolumn{1}{c|}{-1.5} & \multicolumn{1}{c|}{12.8} & \multicolumn{1}{c|}{99.0} & \multicolumn{1}{c|}{91.8} & \multicolumn{1}{c|}{0.7} & 0.5 \\ \cline{2-15} 
\multicolumn{1}{|c|}{} & \multirow{5}{*}{\begin{tabular}[c]{@{}c@{}}Example \ref{example-Separable-density}\\ $t_5$\end{tabular}} & 100 & \multicolumn{1}{c|}{0.2} & \multicolumn{1}{c|}{70.9} & \multicolumn{1}{c|}{99.6} & \multicolumn{1}{c|}{96.4} & \multicolumn{1}{c|}{6.2} & 3.5 & \multicolumn{1}{c|}{-3.6} & \multicolumn{1}{c|}{76.3} & \multicolumn{1}{c|}{98.4} & \multicolumn{1}{c|}{95.4} & \multicolumn{1}{c|}{6.5} & 3.7 \\ \cline{3-15} 
\multicolumn{1}{|c|}{} & & 200 & \multicolumn{1}{c|}{1.1} & \multicolumn{1}{c|}{49.3} & \multicolumn{1}{c|}{99.2} & \multicolumn{1}{c|}{96.6} & \multicolumn{1}{c|}{3.2} & 2.0 & \multicolumn{1}{c|}{4.0} & \multicolumn{1}{c|}{52.1} & \multicolumn{1}{c|}{99.2} & \multicolumn{1}{c|}{96.2} & \multicolumn{1}{c|}{3.3} & 2.0 \\ \cline{3-15} 
\multicolumn{1}{|c|}{} & & 400 & \multicolumn{1}{c|}{0.5} & \multicolumn{1}{c|}{33.1} & \multicolumn{1}{c|}{99.6} & \multicolumn{1}{c|}{97.6} & \multicolumn{1}{c|}{1.8} & 1.3 & \multicolumn{1}{c|}{3.6} & \multicolumn{1}{c|}{33.6} & \multicolumn{1}{c|}{99.4} & \multicolumn{1}{c|}{97.8} & \multicolumn{1}{c|}{1.9} & 1.3 \\ \cline{3-15} 
\multicolumn{1}{|c|}{} & & 800 & \multicolumn{1}{c|}{-0.6} & \multicolumn{1}{c|}{20.7} & \multicolumn{1}{c|}{99.4} & \multicolumn{1}{c|}{97.8} & \multicolumn{1}{c|}{1.1} & 0.9 & \multicolumn{1}{c|}{2.5} & \multicolumn{1}{c|}{22.9} & \multicolumn{1}{c|}{99.4} & \multicolumn{1}{c|}{97.0} & \multicolumn{1}{c|}{1.1} & 0.9 \\ \cline{3-15} 
\multicolumn{1}{|c|}{} & & 1600 & \multicolumn{1}{c|}{-0.2} & \multicolumn{1}{c|}{13.9} & \multicolumn{1}{c|}{98.8} & \multicolumn{1}{c|}{96.6} & \multicolumn{1}{c|}{0.7} & 0.6 & \multicolumn{1}{c|}{0.5} & \multicolumn{1}{c|}{13.7} & \multicolumn{1}{c|}{99.8} & \multicolumn{1}{c|}{98.0} & \multicolumn{1}{c|}{0.7} & 0.6 \\ \cline{2-15} 
\multicolumn{1}{|c|}{} & \multirow{5}{*}{\begin{tabular}[c]{@{}c@{}}Example \ref{example-Isotropic-density}\\ Mat\'ern with $s=7.5$\end{tabular}} & 100 & \multicolumn{1}{c|}{1.2} & \multicolumn{1}{c|}{61.5} & \multicolumn{1}{c|}{98.0} & \multicolumn{1}{c|}{93.4} & \multicolumn{1}{c|}{4.5} & 2.6 & \multicolumn{1}{c|}{-6.5} & \multicolumn{1}{c|}{66.7} & \multicolumn{1}{c|}{95.8} & \multicolumn{1}{c|}{91.2} & \multicolumn{1}{c|}{5.1} & 2.8 \\ \cline{3-15} 
\multicolumn{1}{|c|}{} & & 200 & \multicolumn{1}{c|}{0.7} & \multicolumn{1}{c|}{41.2} & \multicolumn{1}{c|}{97.8} & \multicolumn{1}{c|}{94.4} & \multicolumn{1}{c|}{2.4} & 1.5 & \multicolumn{1}{c|}{0.4} & \multicolumn{1}{c|}{42.5} & \multicolumn{1}{c|}{97.2} & \multicolumn{1}{c|}{93.4} & \multicolumn{1}{c|}{2.5} & 1.6 \\ \cline{3-15} 
\multicolumn{1}{|c|}{} & & 400 & \multicolumn{1}{c|}{1.4} & \multicolumn{1}{c|}{28.3} & \multicolumn{1}{c|}{98.4} & \multicolumn{1}{c|}{94.2} & \multicolumn{1}{c|}{1.5} & 1.0 & \multicolumn{1}{c|}{0.8} & \multicolumn{1}{c|}{28.2} & \multicolumn{1}{c|}{98.6} & \multicolumn{1}{c|}{93.8} & \multicolumn{1}{c|}{1.5} & 1.0 \\ \cline{3-15} 
\multicolumn{1}{|c|}{} & & 800 & \multicolumn{1}{c|}{-1.1} & \multicolumn{1}{c|}{18.2} & \multicolumn{1}{c|}{98.8} & \multicolumn{1}{c|}{96.4} & \multicolumn{1}{c|}{0.9} & 0.7 & \multicolumn{1}{c|}{-0.4} & \multicolumn{1}{c|}{20.1} & \multicolumn{1}{c|}{98.4} & \multicolumn{1}{c|}{94.2} & \multicolumn{1}{c|}{1.0} & 0.7 \\ \cline{3-15} 
\multicolumn{1}{|c|}{} & & 1600 & \multicolumn{1}{c|}{-0.8} & \multicolumn{1}{c|}{13.5} & \multicolumn{1}{c|}{97.6} & \multicolumn{1}{c|}{94.2} & \multicolumn{1}{c|}{0.6} & 0.5 & \multicolumn{1}{c|}{-0.7} & \multicolumn{1}{c|}{13.2} & \multicolumn{1}{c|}{99.4} & \multicolumn{1}{c|}{95.4} & \multicolumn{1}{c|}{0.6} & 0.5 \\ \cline{2-15} 
\multicolumn{1}{|c|}{} & \multirow{5}{*}{\begin{tabular}[c]{@{}c@{}}Example \ref{example-Energy Distance}\\ Energy\end{tabular}} & 100 & \multicolumn{1}{c|}{2.1} & \multicolumn{1}{c|}{62.4} & \multicolumn{1}{c|}{97.8} & \multicolumn{1}{c|}{92.8} & \multicolumn{1}{c|}{4.4} & 2.4 & \multicolumn{1}{c|}{-12.0} & \multicolumn{1}{c|}{68.6} & \multicolumn{1}{c|}{94.6} & \multicolumn{1}{c|}{88.8} & \multicolumn{1}{c|}{4.9} & 2.6 \\ \cline{3-15} 
\multicolumn{1}{|c|}{} & & 200 & \multicolumn{1}{c|}{0.9} & \multicolumn{1}{c|}{39.2} & \multicolumn{1}{c|}{97.6} & \multicolumn{1}{c|}{91.4} & \multicolumn{1}{c|}{2.4} & 1.5 & \multicolumn{1}{c|}{-4.7} & \multicolumn{1}{c|}{42.4} & \multicolumn{1}{c|}{96.0} & \multicolumn{1}{c|}{90.2} & \multicolumn{1}{c|}{2.5} & 1.5 \\ \cline{3-15} 
\multicolumn{1}{|c|}{} & & 400 & \multicolumn{1}{c|}{1.0} & \multicolumn{1}{c|}{25.5} & \multicolumn{1}{c|}{98.8} & \multicolumn{1}{c|}{90.4} & \multicolumn{1}{c|}{1.4} & 0.9 & \multicolumn{1}{c|}{-2.1} & \multicolumn{1}{c|}{27.5} & \multicolumn{1}{c|}{97.6} & \multicolumn{1}{c|}{90.0} & \multicolumn{1}{c|}{1.5} & 0.9 \\ \cline{3-15} 
\multicolumn{1}{|c|}{} & & 800 & \multicolumn{1}{c|}{-0.8} & \multicolumn{1}{c|}{17.2} & \multicolumn{1}{c|}{98.0} & \multicolumn{1}{c|}{89.2} & \multicolumn{1}{c|}{0.9} & 0.6 & \multicolumn{1}{c|}{-1.6} & \multicolumn{1}{c|}{18.3} & \multicolumn{1}{c|}{98.0} & \multicolumn{1}{c|}{88.4} & \multicolumn{1}{c|}{1.0} & 0.6 \\ \cline{3-15} 
\multicolumn{1}{|c|}{} & & 1600 & \multicolumn{1}{c|}{-0.7} & \multicolumn{1}{c|}{11.8} & \multicolumn{1}{c|}{98.0} & \multicolumn{1}{c|}{89.2} & \multicolumn{1}{c|}{0.6} & 0.4 & \multicolumn{1}{c|}{-1.4} & \multicolumn{1}{c|}{11.8} & \multicolumn{1}{c|}{98.8} & \multicolumn{1}{c|}{90.2} & \multicolumn{1}{c|}{0.6} & 0.4 \\ 
\hline
\multicolumn{2}{|c|}{\multirow{5}{*}{IPW}} & 100 & \multicolumn{1}{c|}{8.5} & \multicolumn{1}{c|}{103.3} & \multicolumn{1}{c|}{-} & \multicolumn{1}{c|}{99.0} & \multicolumn{1}{c|}{-} & 7.6 & \multicolumn{1}{c|}{-35.4} & \multicolumn{1}{c|}{80.2} & \multicolumn{1}{c|}{-} & \multicolumn{1}{c|}{95.0} & \multicolumn{1}{c|}{-} & 5.8 \\ \cline{3-15} 
\multicolumn{2}{|c|}{} & 200 & \multicolumn{1}{c|}{0.2} & \multicolumn{1}{c|}{52.4} & \multicolumn{1}{c|}{-} & \multicolumn{1}{c|}{96.0} & \multicolumn{1}{c|}{-} & 2.5 & \multicolumn{1}{c|}{-33.4} & \multicolumn{1}{c|}{53.0} & \multicolumn{1}{c|}{-} & \multicolumn{1}{c|}{91.0} & \multicolumn{1}{c|}{-} & 2.2 \\ \cline{3-15} 
\multicolumn{2}{|c|}{} & 400 & \multicolumn{1}{c|}{0.0} & \multicolumn{1}{c|}{34.2} & \multicolumn{1}{c|}{-} & \multicolumn{1}{c|}{96.0} & \multicolumn{1}{c|}{-} & 1.5 & \multicolumn{1}{c|}{-32.3} & \multicolumn{1}{c|}{38.0} & \multicolumn{1}{c|}{-} & \multicolumn{1}{c|}{83.0} & \multicolumn{1}{c|}{-} & 1.4 \\ \cline{3-15} 
\multicolumn{2}{|c|}{} & 800 & \multicolumn{1}{c|}{-0.3} & \multicolumn{1}{c|}{26.5} & \multicolumn{1}{c|}{-} & \multicolumn{1}{c|}{92.8} & \multicolumn{1}{c|}{-} & 1.0 & \multicolumn{1}{c|}{-33.2} & \multicolumn{1}{c|}{25.7} & \multicolumn{1}{c|}{-} & \multicolumn{1}{c|}{71.6} & \multicolumn{1}{c|}{-} & 1.0 \\ \cline{3-15} 
\multicolumn{2}{|c|}{} & 1600 & \multicolumn{1}{c|}{-0.5} & \multicolumn{1}{c|}{17.5} & \multicolumn{1}{c|}{-} & \multicolumn{1}{c|}{94.6} & \multicolumn{1}{c|}{-} & 0.7 & \multicolumn{1}{c|}{-33.1} & \multicolumn{1}{c|}{17.0} & \multicolumn{1}{c|}{-} & \multicolumn{1}{c|}{52.8} & \multicolumn{1}{c|}{-} & 0.7 \\ \hline
\multicolumn{2}{|c|}{\multirow{5}{*}{CBPS}} & 100 & \multicolumn{1}{c|}{5.4} & \multicolumn{1}{c|}{74.4} & \multicolumn{1}{c|}{-} & \multicolumn{1}{c|}{96.6} & \multicolumn{1}{c|}{-} & 3.7 & \multicolumn{1}{c|}{-31.7} & \multicolumn{1}{c|}{79.8} & \multicolumn{1}{c|}{-} & \multicolumn{1}{c|}{93.2} & \multicolumn{1}{c|}{-} & 3.8 \\ \cline{3-15} 
\multicolumn{2}{|c|}{} & 200 & \multicolumn{1}{c|}{-0.3} & \multicolumn{1}{c|}{50.6} & \multicolumn{1}{c|}{-} & \multicolumn{1}{c|}{95.6} & \multicolumn{1}{c|}{-} & 2.1 & \multicolumn{1}{c|}{-31.0} & \multicolumn{1}{c|}{53.3} & \multicolumn{1}{c|}{-} & \multicolumn{1}{c|}{90.8} & \multicolumn{1}{c|}{-} & 2.2 \\ \cline{3-15} 
\multicolumn{2}{|c|}{} & 400 & \multicolumn{1}{c|}{-0.1} & \multicolumn{1}{c|}{33.4} & \multicolumn{1}{c|}{-} & \multicolumn{1}{c|}{96.2} & \multicolumn{1}{c|}{-} & 1.4 & \multicolumn{1}{c|}{-31.2} & \multicolumn{1}{c|}{38.1} & \multicolumn{1}{c|}{-} & \multicolumn{1}{c|}{84.0} & \multicolumn{1}{c|}{-} & 1.4 \\ \cline{3-15} 
\multicolumn{2}{|c|}{} & 800 & \multicolumn{1}{c|}{-0.2} & \multicolumn{1}{c|}{26.2} & \multicolumn{1}{c|}{-} & \multicolumn{1}{c|}{92.6} & \multicolumn{1}{c|}{-} & 1.0 & \multicolumn{1}{c|}{-32.6} & \multicolumn{1}{c|}{25.7} & \multicolumn{1}{c|}{-} & \multicolumn{1}{c|}{73.6} & \multicolumn{1}{c|}{-} & 1.0 \\ \cline{3-15} 
\multicolumn{2}{|c|}{} & 1600 & \multicolumn{1}{c|}{-0.5} & \multicolumn{1}{c|}{17.4} & \multicolumn{1}{c|}{-} & \multicolumn{1}{c|}{94.6} & \multicolumn{1}{c|}{-} & 0.7 & \multicolumn{1}{c|}{-32.8} & \multicolumn{1}{c|}{17.0} & \multicolumn{1}{c|}{-} & \multicolumn{1}{c|}{54.4} & \multicolumn{1}{c|}{-} & 0.7 \\ \hline
\end{tabular}

\caption{\footnotesize Summary of the simulation studies. Bias and ESE are scaled by a factor of 100.}
\label{tab:simulation} 
\end{table}}%

\section{Application}\label{sec-real_data}

We apply the proposed method to a real-world setting, following the analysis of \citet{DDML2018}, and focus on the effect of 401(k) participation on financial asset accumulation. 
We use data from the \texttt{DoubleML} R package \citep{DoubleML2022Python}, which includes $n = 9,915$ individuals, of whom $3,682$ are eligible for a 401(k) plan. The primary variables are net financial assets ($Y$), 401(k) participation ($A$), and 401(k) eligibility ($Z$). To adjust for confounding, we include nine pre-treatment covariates ($\bX$): four continuous variables (age, income, family size, and years of education) and five binary variables (defined benefit pension status, marital status, two-earner household, IRA participation, and home ownership). All continuous covariates are standardized to the $[0,1]$ interval prior to analysis. 

Due to potential unmeasured confounding between 401(k) participation and net financial assets, we focus on estimating the LATE following the approach in Section \ref{sec-IV}, rather than the overall ATE. Specifically, in this application, the LATE is defined as the effect of 401(k) participation on net financial assets for individuals who would participate in 401(k) if eligible, but not otherwise. The analysis relies on 401(k) eligibility serving as a valid IV, satisfying Assumption \ref{assumption IV}; for further details, we refer readers to Section 6.2 of \citet{DDML2018}.

We compare eleven estimators in our analysis. First, we include the same five weighting estimators examined in the simulation study in Section \ref{sec-simulation}, namely: (i) Gaussian, (ii) $\ell_1$-Laplacian, (iii) $t_5$, (iv) Mat\'ern, and (v) energy. Their implementation remains largely unchanged, employing three-way balancing, the median heuristic for hyperparameter tuning, and a regularization parameter of $\lambda = n^{-2}$. The only modification is that the smoothness parameter of the Mat\'ern kernel is set to $s = 4.5$ in order to account for the smaller number of continuous covariates. For these five estimators, we report 95\% confidence intervals based on both subsampling and bootstrap methods. Second, we consider the same two IPW and CBPS estimators with identical specifications as in the simulation study, reporting 95\% bootstrap confidence intervals. Finally, we report four double/debiased machine learning (DDML) estimators, each based on a different nonparametric approach, which are directly available from the online repository of \citet{DoubleML2022Python}. DDML differs from weighting-based approaches in that it fits flexible nuisance regressions (e.g., treatment and outcome models) and then applies an Neyman orthogonal (influence function-based) estimating equation with cross-fitting to enable valid statistical inference. We include these DDML results as additional benchmarks in this canonical application, even though our simulation study focuses on weighting-based methods.


The results of the data analysis are summarized in Table \ref{tab-data}. We first focus on the LATE estimates obtained from our five weighting estimators. The point estimates are largely consistent across different density specifications, with values around \$12,000. Considering the 95\% confidence intervals, we further conclude that the LATE is significantly different from zero at the 5\% significance level, indicating that 401(k) participation increases net financial assets. Consistent with our simulation findings, the subsampling confidence intervals are wider than the corresponding bootstrap intervals. However, as noted above, the standard bootstrap may fail to achieve the nominal coverage level for our estimator. Therefore, despite being slightly more conservative, we recommend reporting the subsampling confidence intervals as the ultimate measure of uncertainty. 

We now turn to a comparison between our method and alternative approaches. Notably, our estimates are closely aligned with those produced by the DDML methods. This agreement with a widely used nonparametric approach further reinforces the credibility of our estimator. In contrast, the point estimates from the IPW and CBPS methods differ substantially, likely due to misspecification in the assumed parametric models. Regarding the confidence intervals, our weighting estimator produces slightly wider intervals (based on subsampling) than the DDML estimator. This modest loss in efficiency may be attributed to two factors: (i) our method does not incorporate outcome information, and (ii) the DDML estimator may attain the semiparametric efficiency bound under mild regularity conditions \citep{DDML2018}. We anticipate that the augmented weighting estimator discussed in Remark \ref{augmentation} will yield confidence intervals of comparable length. Overall, these empirical findings indicate that our proposed method performs competitively relative to state-of-the-art approaches in this real-world application.

\begin{table}[!htp]
\renewcommand{\arraystretch}{1.1}
\scriptsize
\centering
\setlength{\tabcolsep}{8pt}
\begin{tabular}{|cc|c|c|c|}
\hline
\multicolumn{2}{|c|}{Method} & Estimate & 95\% CI & 95\% CI Length \\ \hline
\multicolumn{1}{|c|}{\multirow{10}{*}{\begin{tabular}[c]{@{}c@{}}Distributional\\ Balancing\end{tabular}}} & \multirow{2}{*}{\begin{tabular}[c]{@{}c@{}}Example \ref{example-Gaussian density}\\ Gaussian\end{tabular}} & \multirow{2}{*}{11927} & $\makebox[1cm][l]{\text{SS}:}$ (7153, 17140) & 9988 \\ \cline{4-5} 
\multicolumn{1}{|c|}{} & & & $\makebox[1cm][l]{\text{Boot}:}$ (7405, 14792) & 7387 \\ \cline{2-5} 
\multicolumn{1}{|c|}{} & \multirow{2}{*}{\begin{tabular}[c]{@{}c@{}}Example \ref{example-Separable-density}\\ $\ell_1$-Laplacian\end{tabular}} & \multirow{2}{*}{12486} & $\makebox[1cm][l]{\text{SS}:}$ (8984, 16625) & 7641 \\ \cline{4-5} 
\multicolumn{1}{|c|}{} & & & $\makebox[1cm][l]{\text{Boot}:}$ (8330, 15076) & 6746 \\ \cline{2-5} 
\multicolumn{1}{|c|}{} & \multirow{2}{*}{\begin{tabular}[c]{@{}c@{}}Example \ref{example-Separable-density}\\ $t_5$\end{tabular}} & \multirow{2}{*}{12828} & $\makebox[1cm][l]{\text{SS}:}$ (7813, 17641) & 9828 \\ \cline{4-5} 
\multicolumn{1}{|c|}{} & & & $\makebox[1cm][l]{\text{Boot}:}$ (8135, 15126) & 6990 \\ \cline{2-5} 
\multicolumn{1}{|c|}{} & \multirow{2}{*}{\begin{tabular}[c]{@{}c@{}}Example \ref{example-Isotropic-density}\\ Mat\'ern with $s=4.5$\end{tabular}} & \multirow{2}{*}{12307} & $\makebox[1cm][l]{\text{SS}:}$ (7613, 17234) & 9621 \\ \cline{4-5} 
\multicolumn{1}{|c|}{} & & & $\makebox[1cm][l]{\text{Boot}:}$ (7887, 15251) & 7364 \\ \cline{2-5} 
\multicolumn{1}{|c|}{} & \multirow{2}{*}{\begin{tabular}[c]{@{}c@{}}Example \ref{example-Energy Distance}\\ Energy\end{tabular}} & \multirow{2}{*}{12619} & $\makebox[1cm][l]{\text{SS}:}$ (9399, 16802) & 7404 \\ \cline{4-5} 
\multicolumn{1}{|c|}{} & & & $\makebox[1cm][l]{\text{Boot}:}$ (8206, 14359) & 6153 \\ \hline
\multicolumn{2}{|c|}{IPW} & 444 & $\makebox[1cm][l]{}$ (-12648, 9684) & 22333 \\ \hline
\multicolumn{2}{|c|}{CBPS} & 5769 & $\makebox[1cm][l]{}$ (-631, 12024) & 12655 \\ \hline
\multicolumn{1}{|c|}{\multirow{4}{*}{DDML}} & glmnet & 12802 & $\makebox[1cm][l]{}$ (8999, 16606) & 7607 \\ \cline{2-5} 
\multicolumn{1}{|c|}{} & ranger & 11792 & $\makebox[1cm][l]{}$ (8649, 14936) & 6287 \\ \cline{2-5} 
\multicolumn{1}{|c|}{} & rpart & 12214 & $\makebox[1cm][l]{}$ (8856, 15573) & 6717 \\ \cline{2-5} 
\multicolumn{1}{|c|}{} & xbgoost & 11861 & $\makebox[1cm][l]{}$ (8687, 15035) & 6348 \\ \hline
\end{tabular}
\caption{\footnotesize Summary of the 401(k) dataset analysis. All figures are presented in U.S. dollars.} 
\label{tab-data}
\end{table}

\section{Discussion}\label{sec:discussion}

In this paper, we introduced a unified nonparametric framework for estimating causal effects using CFD-based balancing weights. In particular, we showed that some existing methods can be viewed as special cases of the proposed framework. Our theoretical results establish the $\sqrt{n}$-consistency of the estimator under suitable regularity conditions. We further highlight several aspects that are often overlooked in practice. First, we clarify the connection between the specification of the density function (or the RKHS kernel) and the corresponding requirements on the outcome regression (see Assumption \ref{assumption-mu}). Second, we provide both theoretical justification and simulation evidence showing that the standard bootstrap may fail, while subsampling offers a valid alternative. We extend our approach to an IV setting to account for potential unmeasured confounding, demonstrating the broad applicability of our method. Simulation and empirical analyses under this IV setting demonstrate that the proposed estimator is robust to model misspecification, in contrast to standard parametric alternatives, and performs competitively with the DDML estimator. 


We conclude the paper by outlining potential directions for future research. First, the proposed CFD-based method requires the investigator to specify the density $\omega$ and its associated hyperparameters. We have not addressed how to select these quantities optimally based on the performance of the resulting weighting estimator. Developing a data-driven cross-validation procedure for selecting $\omega$ and its hyperparameters by directly optimizing the causal effect estimator remains an important open direction. Second, the proposed estimator requires solving a constrained QP involving a gram matrix of size $n \times n$. For large $n$, this computation can be burdensome, motivating the development of more efficient algorithms for exceptionally large datasets. We believe that approximation techniques (e.g., Nystr\"om method or sampling-based approaches) are promising directions for future research. Finally, in Section \ref{sec-simulation}, we find that our subsampling confidence intervals are generally conservative. Future research could explore ways to refine subsampling procedures to achieve nominal coverage without being overly conservative, or to develop inferential methods that are theoretically valid for estimators derived from constrained optimization problems.


\noindent \textbf{Supplementary Materials}:
The supplementary materials contain the proofs of theoretical results, additional numerical results, and R code and data.

\noindent \textbf{Data Availability Statement}:
The data that support the findings of this study are openly available in DoubleML at \url{https://docs.doubleml.org/stable/index.html}.

\def\spacingset#1{\renewcommand{\baselinestretch}%
{#1}\small\normalsize} \spacingset{1.5}

\bibliographystyle{apa}
\bibliography{reference_diptanil.bib}

\end{document}